 \documentclass[final,5p,times,twocolumn,authoryear]{elsarticle}

\usepackage{amssymb}
\usepackage{lipsum}
\usepackage{xcolor}
\usepackage[colorlinks=true, linkcolor=blue, urlcolor=blue, citecolor=blue]{hyperref}
\usepackage{graphicx}
\usepackage{caption}	% Including figure files
\usepackage{subcaption}	% Including figure files
\usepackage{float}
\usepackage{hyperref}
\usepackage{amsmath}
\usepackage{multirow}
\usepackage{lineno}
\def\mnras{MNRAS}
\def\apj{ApJ}
\def\nat{Nature}

\def\apjl{ApJL}
\def\apjs{ApJS}
\def\aap{A\&A}
\def\aj{AJ}

\def\pasj{PASJ}

\def\prl{PRL}
\def\physrep{PHYSREP}
\def\prd{PRD}
\def\nar{NAR}
\def\textasciitildeJ{tldaJ}
\newcommand{\nicer}{\textit{NICER}}
\newcommand{\xmm}{\textit{XMM-Newton}}

\newcommand{\nus}{\textit{NuSTAR}}
\newcommand{\chandra}{\textit{Chandra}}
\journal{High Energy Astrophysics}

\begin{document}

\begin{frontmatter}



\title{Measuring accretion disc properties in the transitional millisecond pulsar PSR J1023+0038 using \xmm{}, \nus{}, \nicer{} and \chandra{}}

%\author[label1,label2]{}

\author[a]{Vishal Jadoliya}
\affiliation[a]{organization={Department of Physics, Indian Institute of Technology Hyderabad},%Department and Organization
            addressline={IITH main road, Kandi, Sangareddy}, 
            city={Hyderabad},
            postcode={502284}, 
            state={Telangana},
            country={India}}

            \author[a]{Mayukh Pahari}

\author[b,c]{Sudip Bhattacharyya}
\affiliation[b]{organization={Department of Astronomy and Astrophysics, Tata Institute of Fundamental Research},%Department and Organization
            addressline={1 Homi Bhabha Road, Colaba}, 
            city={Mumbai},
            postcode={400005}, 
            state={Maharashtra},
            country={India}}
\affiliation[c]{organization={MIT Kavli Institute for Astrophysics and Space Research, 
Massachusetts Institute of Technology},%Department and Organization
            addressline={70 Vassar St}, 
            city={Cambridge},
            state={MA},
            postcode={02139}, 
            country={USA}}
\author[d]{Shaswat Suresh Nair}
\affiliation[d]{organization={Indian Institute of Science Education and Research, Pune},%Department and Organization
            addressline={IISER Pune, Dr. Homi Bhabha Road}, 
            city={Pune},
            postcode={411008}, 
            state={Maharashtra},
            country={India}}
%\affiliation[c]{organization={Department of Physics, Indian Institute of Technology Hyderabad},%Department and Organization
            %addressline={IITH main road, Kandi, Sangareddy}, 
            %city={Hyderabad},
            %postcode={502284}, 
            %state={Telangana},
            %country={India}}
\begin{abstract}
Whether the accretion disc in the X-ray high-mode of transitional millisecond pulsars (tMSP) reaches near the neutron star surface by penetrating the magnetosphere is a crucial question with many implications, including for continuous gravitational wave emission from the pulsar. 
We attempt to answer this question for the tMSP PSR J1023+0038 by segregating high-mode data and performing detailed spectral analysis using the \xmm{} EPIC-PN+MOS1+MOS2 joint observations, \xmm{}+\nus{} joint observations, \nicer{} and \chandra{} individual observations during different epochs. 
With the sum of longest exposures ($\sim$202 ksec of high mode data from $\sim$364 ksec of total exposure), we performed a self-consistent spectral analysis and constrain the inner disc radius 16.8 $\pm$ 3.8 km with at least 3$\sigma$ significance. 
Such a measurement is found consistent with best-fit spectral values of inner disc radius from other observatory like \nicer{} and a joint observations with \xmm{} and \nus{} within 3$\sigma$ limits.
We also detect a Fe emission line at 6.45 keV, for the first time from a tMSP, in the \chandra{} spectrum with 99\% significance with an upper limit of the inner disc radius of 21 R$_g$, supporting independently the fact that inner disc extends into neutron stars's magnetosphere during high mode. 
All results from our analysis imply that the accretion disc is significantly present and extended within the corotation radius of the neutron star in PSR J1023+0038 during the X-ray high-mode of the tMSP PSR J1023+0038.
The measured range of inner disc radius is fully consistent with an independent analysis by \citet{Bhattacharyya2020}, which suggests continuous gravitational wave emission from this neutron star, and the standard model of X-ray pulsations in accreting MSPs.
\end{abstract}


\end{frontmatter}

\section{Introduction}
\label{introduction}

An accreting neutron star (NS) in a low-mass X-ray binary system (LMXBs) can evolve into a radio millisecond pulsar (radio MSP) at its final stage of evolution through a prolonged accretion of matter from a low-mass companion star (e.g., a low-mass main sequence star, a white dwarf, etc.). The MSPs are generally old pulsars and their short spin periods ($<$ 30 ms) are well explained by the recycling scenario first proposed in the 1980s, in which the NS spins up due to long phase ($\sim$ Gyr) of sustained mass and angular momentum transfer via an accretion disc in a LMXBs. \citep{Alpar1982,Radhakrishnan1982,Fabian1983,Taam1986,Bhattacharya1991,Tauris2006,Srinivasan2010,Papitto2022}. Over the past decade, observations of some MSPs have revealed that these systems can undergo a dramatic and rapid state transition on a time scale of a few weeks. Specifically, the transition occurs between a disc-dominated accretion-powered millisecond X-ray pulsar (AMXP) state exhibiting the emission characteristics of LMXBs and a disc-free rotation-powered MSP state where they behave like a radio millisecond pulsar (RMSP) \citep{Archibald2009,Papitto2013,Bassa2014}. 
This new class of MSPs became known as transitional millisecond pulsars (tMSPs) 
\citep{Papitto2022}.

Among the three confirmed tMSPs---IGR J18245-2452 \citep{Papitto2013}, PSR J1023+0038 \citep{Archibald2009}, and XSS J12270-4859 \citep{Bassa2014}---PSR J1023+0038 (also known as FIRST J102347.6+003841 or AY Sextantis) is one of the most well-studied system not only because of its proximity \citep[$\rm 1.37\pm0.04 \, kpc$;][]{Deller2012} but also because it shows many unique aspects. For example, so far this is the only MSP for which the rate of change of the NS spin frequency \citep[$\rm \nu \simeq 592.42 \,Hz$;][]{Archibald2009,Deller2012} has been measured in both AMXP and RMSP states \citep[$\rm \dot\nu_{AMXP} \sim -2.50\times10^{-15}\, Hz\,s^{-1}$; $\rm \dot\nu_{RMSP} \sim -1.89\times10^{-15}\, Hz\,s^{-1}$][]{Deller2012,Archibald2013,Jaodand2016,Papitto2022}.
This provides a unique opportunity to infer the emission of continuous gravitational waves from a slightly deformed neutron star by identifying the contribution of these waves to the $\dot\nu$ values \citep{Haskell2017,Bhattacharyya2020}.

The tMSP PSR J1023+0038 was observed as an eclipsing radio pulsar that has an NS of mass $\rm 1.71\pm0.16 \, M_{\odot}$ with a pulse period of 1.69 ms \citep{Archibald2009,Deller2012}, along with an orbital period of $4.75$~h around a Roche lobe that fills a G6-type low-mass companion star of mass $\rm \simeq 0.2 \, M_{\odot}$ \citep{Thorstensen2005,Archibald2010,Tendulkar2014}. On June 23, 2013, observations showed a dramatic cessation in the radio pulsations of PSR J1023+0038, and this phase has continued to date. Subsequent multi-wavelength (X-ray, $\gamma$-ray, and optical) observations revealed an increase in flux, indicating that PSR J1023+0038 had undergone a state transition from a RMSP state to a peculiar AMXP state, characterised by the accretion disc dominated low-luminosity emission with X-ray luminosity ($L_{\rm X}$) of $\sim \rm 10^{33-34} erg\,s^{-1}$, referred to as X-ray sub-luminous state or active state \citep{Halpern2013,Patruno2014,Stappers2014,Tendulkar2014,Campana2018}. 
Moreover, in the AMXP state, the $\gamma$-ray luminosity ($ L^{\rm AMXP}_\gamma$) of PSR J1023+0038 was observed to be $\rm (6.0-7.8)\times 10^{33}\,erg\,s^{-1}$ by \citet{Stappers2014,Deller2015}; however, \citet{Torres2017} reported a higher value of $\rm L^{\rm AMXP}_\gamma \approx 12.4\times 10^{33}\,erg\,s^{-1}$. In comparison, during the RMSP state, the $\gamma$-ray luminosity ($ L^{\rm RMSP}_\gamma$) was lower by a factor of $\rm \sim 5-6.5$ relative to the AMXP state, with an observed value of $\rm \sim 1.2\times 10^{33}\,erg\,s^{-1}$ \citep{Nolan2012,Papitto2015}.
%%%%%%%%%%%%%%%%%%%%%%%%%%%%%%%%%%%%%%%%%%%%%%%%%%%%%%%%%%%%%%%%%%%%%%%%%%%%%%%%%%%%%%%%%%%%%%%%%%%%%%%%%%%%%%%%%%%%%%%%%%%%%%%%%%%%%%%%%%%%%%%%%%%%%%%%%%%%%%%%%%%%%%%%%%%%%%%%%%%%%%%%%%%%%%%%%%%%%%%%%%%%%%

\begin{table*}[!ht]
\setlength{\tabcolsep}{4pt} 
\centering
\caption{Log of observations of PSR J1023+0038.}
\renewcommand{\arraystretch}{1.2} 
\label{tab:log}
\begin{tabular}{lcccccccccr} 
\hline
\hline
Observatory    & Observation     & Observation Date  & Start time   & Instrument      & Exposure  &  High-mode \\
             & ID              & (YYYY-MM-DD)      & UT(hh:mm:ss) & (Mode)          & Time (ks) &  duration (ks)  \\
\hline
{\it XMM-Newton} & 0720030101(X1)  & 2013-11-10        & 15:42:34     & EPIC MOS (SW)   & 129.2            & 90.1     \\
             &                 &                   &              & EPIC PN (FT)    & 128.4            & 91.6     \\
             & 0742610101(X2)  & 2014-06-10        & 03:27:28     & EPIC MOS (SW)   & 118.5            & 60.4     \\
             &                 &                   &              & EPIC PN (FT)    & 116.7            & 61.3    \\
             & 0784700201(X3)  & 2016-05-08        & 03:29:38     & EPIC MOS (SW)   & 116.4            & 48.3     \\
             &                 &                   &              & EPIC PN (FT)    & 123.9            & 49.0     \\
             & 0864010101(X4)  & 2021-06-03        & 19:48:47     & EPIC MOS (SW)   & 59.4             & 39.5     \\
             &                 &                   &              & EPIC PN (FT)    & 60.9             & 39.8     \\
\hline
{\it NuSTAR} &30201005002(N1)  & 2016-05-07        & 15:21:10     & FPMA (Science)  & 87.3             & 47.9      \\
             &                 &                   &              & FPMB (Science)  & 87.2             & 47.8       \\
             &30601005002(N2)  & 2021-06-03        & 20:36:06     & FPMA (Science)  & 22.5             & 13.5       \\
             &                 &                   &              & FPMB (Science)  & 22.3             & 13.4       \\             
\hline
\nicer{}     & 2531010101      & 2019-05-08        & 00:37:40     & XTI             & 29.0             & 22.0        \\
             & 3515010101      & 2020-04-10        & 02:47:32     & XTI             & 28.2             & 20.5        \\
\hline
{\it Chandra}& 17785           & 2016-10-21        & 10:41:28     & ACIS-S (CC)     & 20.2             & 13.7          \\
\hline
\end{tabular}
\begin{flushleft}
\textbf{Note:}  X3,N1 and X4,N2 are Simultaneous observations with {\it XMM-Newton}. X2, X3, and X4 observations of {\it XMM-Newton} are affected by background proton flares.
\end{flushleft}
\end{table*}

%%%%%%%%%%%%%%%%%%%%%%%%%%%%%%%%%%%%%%%%%%%%%%%%%%%%%%%%%%%%%%%%%%%%%%%%%%%%%%%%%%%%%%%%%%%%%%%%%%%%%%%%%%%%%%%%%%%%%%%%%%%%%%%%%%%%%%%%%%%%%%%%%%%%%%%%%%%%%%%%%%%%%%%%%%%%%%%%%%%%%%%%%%%%%%%%%%%%%%%%%
During the active state, PSR J1023+0038 exhibits rapid transitions between three distinct and unique X-ray luminosity modes: high-mode, low-mode, and flaring-mode \citep{Linares2014,Bogdanov2015,Papitto2019}. The source remains in high-mode most of the time ($\sim 77\%$), having a relatively steady X-ray luminosity ($L_X \sim  7 \times 10^{33}\, \rm erg\,s^{-1}, 0.3-79~keV$), which is often accompanied with coherent pulsations (rms amplitude $\sim5-10\%$) at the spin period of the pulsar. 
In contrast, the low-mode occurs for a shorter time ($\sim 22\%$) and exhibits a sudden drop in X-ray luminosity ($\sim 10^{33}\,\rm erg\,s^{-1}, 0.3-79~keV$), accompanied by the disappearance of coherent pulsations. The system occasionally ($\sim 1\%$) enters the flaring-mode, marked by brief, irregular bursts of X-ray emission reaching luminosities as high as $\sim 2\times10^{34} \,\rm erg\,s^{-1}$ with no coherent pulsation \citep{Archibald2009,Archibald2010,Archibald2015,Papitto2015,Bogdanov2015,Campana2016,Jaodand2016}. PSR J1023+0038 switches between the respective modes within a timespan of $\sim 10-30$~s in X-rays, while staying in low-mode for a few tens of seconds to minutes \citep{Bogdanov2015}.

The aforementioned observations pose several puzzles. For example, it is challenging to explain the source of the $\gamma$-ray emission in the AMXP state, particularly the several times higher values of this emission compared to the non-accreting RMSP state. Two distinct modes in X-rays in the AMXP state, along with the detection of X-ray pulsations only in the high-mode, are another puzzling feature. 
According to the standard model, accretion-powered pulsations happen when the disc is stopped by the neutron star's magnetosphere at the magnetospheric radius ($r_{\rm m}$), which is less than the corotation radius \citep[$r_{\rm co}$; ][and references therein]{DiSalvo2022}.
The value of $r_{\rm co}$ for this source is $\approx 24.6-26.2 \,{\rm km}$, for $M$ \,$\approx 1.55-1.87 {\rm M_{\odot}}$ \citep{Bhattacharyya2020}. However, if one estimates the pulsar magnetic field from the $\dot\nu_{\rm RMSP}$ then the low accretion rate inferred from a low observed luminosity indicates $r_{\rm m} > r_{\rm co}$ during the high-mode, which does not satisfy the standard condition for X-ray pulsations \citep[][and references therein]{Bhattacharyya2020}.
However, \cite{Bhattacharyya2020}, starting from general torque budget and energy budget equations, inferred that the magnetic field value ($B\approx\,(0.7-6.1)\times\,10^7$~G) of the pulsar is lower than the previously estimated value of $\sim 10^8$~G, and hence the condition $r_{\rm m} < r_{\rm co}$ for standard model of X-ray pulsations could be satisfied for the high-mode.
\cite{Bhattacharyya2020} also inferred an ellipticity of $\sim 10^9$ of the pulsar, which is not unexpected \citep{Johnson-McDaniel2013,Haskell2017,Woan2018}. 
The pulsar should emit gravitational waves continuously due to such an ellipticity.
Note that earlier X-ray spectral analysis also suggested that the $r_{\rm m} < r_{\rm co}$ condition in the high-mode could not be ruled out \citep{Campana2016, CotiZelati2018}. However, the discovery of optical pulsations \citep{Ambrosino2017,Papitto2019,Zampieri2019} in PSR J1023+0038 motivated some authors to propose an alternative model for the high-mode, where the accretion disc is beyond the light cylinder of the pulsar, X-ray pulsations are explained in a non-standard way, and the X-ray spectral models are different \citep{Campana2019,Papitto2019,Veledina2019}. Such a model was proposed despite similar optical/UV pulsations being observed in accreting MSP SAX J1808.4–3658 \citep{Ambrosino2021}, where accretion onto the pulsar is established \citep{Sharma2023,Casten2023,Ballocco2025}.
Thus, for the X-ray high-mode, there are two primary models, for one the disc extends inside the $r_{\rm co}$ and for the other the disc is beyond the light cylinder. It is crucial to find which one is correct because that will have implications for the physics of the neutron star, binary system, accretion, and continuous gravitational wave emission, and will be useful to understand the unique properties of transitional pulsars. In this work, we address this question by investigating multi-mission (\xmm{}, \chandra{}, \nus{}, and \nicer{}) and multi-epoch X-ray observations. 

The paper is organised as follows: 
In Section \ref{multimission_analysis}, we describe X-ray observations with four instruments and the corresponding data reduction.
Section \ref{Results} details our analysis and results.
We discuss implications of the results and make conclusions in Section \ref{Discussion}.

\section{The multi-mission X-ray observations, and data reduction }
\label{multimission_analysis}
We used multi-epoch archival observations of PSR J1023+0038 obtained from \xmm{}, \nus{}, \nicer{}, and \chandra{} X-ray missions. A detailed log of these observations is provided in Table \ref{tab:log}. This section provides an overview of the observations, and the procedures employed for data reduction.

\subsection{XMM-Newton}
We selected and analyzed a set of XMM-Newton observations of PSR J1023+0038 conducted during its active state, spanning the period from 2013 November 10 to 2021 June 3. To ensure robust temporal and spectral analysis, we restricted our sample to high signal-to-noise ratio observations with exposure times exceeding 60 ks.
\subsubsection{XMM-Newton observations and data reduction}
\label{xmm_observation & data reduction}
Observations of PSR J1023+0038 were carried out with the European Photon Imaging Camera (EPIC) instrument on board XMM-Newton, which comprises three co-aligned X-ray telescopes equipped with two MOS \citep{Turner2001} and one PN \citep{Struder2001} CCD. For all the observations, both MOS detectors were configured for small window (SW) mode with a time resolution of 0.3 sec to mitigate the pile-up effect. In contrast, the PN detector was configured for fast timing (FT) mode, providing a higher time resolution of 29.52 ${\mu}s$ with one-dimensional imaging capability.
Data reduction for all data sets was performed with standard \xmm{} Science Analysis Software (SAS) version xmmsas\_22.0.0-9173c7d25-20250128 using the latest Current Calibration Files (CCFs). The EPIC--MOS and EPIC--PN Observation Data Files (ODFs) were locally reprocessed by executing the SAS tasks \textsc{emproc} and \textsc{epproc}, respectively. All observations were checked for high background proton flaring events, and affected ones were filtered out using the standard procedure (See Table \ref{tab:log}). The events files were further filtered with \textsc{xmmselect/evselect} in the energy range 0.3--10.0 keV using the filtering criteria FLAG==0, PATTERN 0--12 (0--4), and \#XMMEA\_EM(P) options for EPIC--MOS(PN). The task \textsc{epatplot} inspected the pileup effect, but none of the observations were affected by the pileup. 

The source events for MOS were extracted from a circular region of 36 arcsec, while PN events were extracted from a region with a width of 7 pixels, except for ObsID 0864010101, where a region of 10 pixels centred on the source was used to ensure optimal source coverage. The background events were extracted from similar source-free regions close to the source in the same CCD of the detector. For each observation, the SAS task \textsc{epiclccorr} was used to obtain the background-subtracted and exposure-corrected light curves for both MOS and PN detectors, binned at a time resolution of 10s. These light curves were then merged using the \textsc{lcmath} task of the \textsc{FTOOLS} package, and all of them exhibited characteristic stable count rates in low, high, and flaring modes (see Figure \ref{fig1:xmm_long_lc}), as reported by previous studies \citep{Bogdanov2015,Archibald2015,Jaodand2016}. Therefore, we generated Good Time Intervals (GTI) using the SAS task \textsc{tabgtigen}, following the high-mode count rate criterion of 4.1--11.0 $\rm counts\,s^{-1}$ (See top left panel of Figure \ref{fig2:xmm_long_spec}), as reported in \citet{Bogdanov2015}. Subsequently, the task \textsc{especget} was executed to extract the source and background spectra along with the corresponding source-specific Ancillary Response Files (ARF) and Redistribution Matrix Files (RMF) using the high-mode GTI. EPIC MOS1 and MOS2 spectra and their respective response files were merged for each observation using the SAS task \textsc{epicspeccombine}, rebinned to have a minimum of 100 counts per spectral bin, and were fitted in the energy range of 0.3--10.0 keV. To obtain good signal-to-noise, EPIC PN spectra were rebinned to have a minimum of 200 counts using the task \textsc{grppha} and fitted in the 0.6--10.0 keV energy range. Furthermore, the merged MOS (MOS1+MOS2) and PN spectra, along with their respective response files from the three longest observations (i.e., ObsID: 0720030101, 0742610101, and 0784700201), were further combined using the same task to produce grand MOS and PN spectra and responses.

%%%%%%%%%%%%%%%%%%%%%%%%%%%%%%%%%%%%%%%%%%%%%%%%%%%%%%%%%%%%%%%%%%%%%%%%%%%%%%%%%%%%%%%%%%%%%%%%%%%%%%%%%%%%%%%%%%%%%%%%%%%%%%%%%%%%%%%%%%%%%%%%%%%%%%%%%%%%%%%%%%%%%%%%%%%%%%%%%%%%%%%%%
\subsection{NuSTAR}
To perform a thorough investigation, a multi-mission approach was adopted by selecting only those \nus{} (Nuclear Spectroscopic Telescope Array) observations that were strictly simultaneous with \xmm{}. PSR J1023+0038 was observed concurrently by NuSTAR and \xmm{} during two observational campaigns conducted on May 2016 and June 2021.

\subsubsection{\nus{} observations and data reduction}
Simultaneous observations of PSR J1023+0038 were carried out with the \nus{}, the first focusing hard X-ray observatory operating in the 3–79 keV energy range. \nus{} consists of two co-aligned grazing-incidence telescopes, designated FPMA and FPMB, each equipped with CdZnTe pixel detectors \citep{Harrison2013}. All observations were performed in standard SCIENCE mode.

Data reduction for both \nus{} observations of PSR J1023+0038 (details provided in Table \ref{tab:log}) was performed using the standard \nus{} Data Analysis Software \textsc{NUSTARDAS version 2.1.4}, distributed as part of \textsc{HEASoft version 6.34}, with the \textsc{NuSTAR CALDB version 20240325}. The raw data from both observations were reprocessed locally using the task \textsc{nupipeline}, which generated cleaned and calibrated event files for the FPMA and FPMB detectors. For each observation, source events from both detectors were extracted using a circular region of 50 arcsec radius, centred on the source. The background events were extracted from a nearby, source-free circular region of the same radius. The task \textsc{nuproducts} was used to generate the source and background light curves and spectra in the 3.0--79.0 keV energy band. Background-subtracted light curves were extracted using the task \textsc{lcmath}, with a time binning of 50 seconds. For mode selection, the light curves from both detectors were combined using \textsc{lcmath} to produce a merged light curve with improved signal-to-noise ratio. Good Time Intervals (GTIs) corresponding to the high-mode were generated using tools from the FTOOL package, based on a count rate threshold criterion of 0.40--1.5 $\rm counts\,s^{-1}$ (See Figure \ref{fig4:nustar_mode_selection}). Subsequently, the task \textsc{nuproducts} was used to extract source and background spectra for the high-mode along with corresponding source-specific ARF and RMF using the high-mode GTI. For each instrument, spectra were rebinned to have a minimum of 20 counts per energy bin using the task \textsc{grppha}.

%%%%%%%%%%%%%%%%%%%%%%%%%%%%%%%%%%%%%%%%%%%%%%%%%%%%%%%%%%%%%%%%%%%%%%%%%%%%%%%%%%%%%%%%%%%%%%%%%%%%%%%%%%%%%%%%%%%%%%%%%%%%%%%%%%%%%%%%%%%%%%%%%%%%%%%%%%%%%%%%%%%%%%%%%%%%%%%%%%%%%%%%%%%
\subsection{\nicer{}}
PSR J1023+0038 has been observed on multiple occasions by the Neutron Star Interior Composition Explorer (\nicer{}) as part of ongoing monitoring campaigns. For this study, we selected and analysed the two longest available exposures, obtained in April 2020 and May 2019, to ensure sufficient photon statistics for detailed temporal and spectral analysis.

\subsubsection{\nicer{} observations and data reduction}
\nicer{} is a non-imaging X-ray observatory mounted on the International Space Station (ISS), optimised for observations in the 0.2–12.0 keV energy band \citep{Keith2012}.  Although \nicer{} lacks imaging capability, it offers moderate spectral resolution of (E\,/\,$\Delta$E) $\sim$85 eV (FWHM) at 1.0 keV. 
Data reduction for the \nicer{} observation of PSR J1023+0038 was carried out using the standard \nicer{} Data Analysis Software \textsc{NICERDAS version 13}, distributed as part of \textsc{HEASoft version 6.34}, with \textsc{NICER CALDB version xti20240206}. The raw, unfiltered observation data were reprocessed locally using the pipeline task \textsc{nicerl2}, which produced cleaned and calibrated Level-2 event files for the X-ray Timing Instrument (XTI). Time series products were extracted in the 0.5–10.0 keV energy band with a time binning of 10 seconds using the task \textsc{nicerl3-lc}. Good Time Intervals (GTIs) corresponding to the high-mode were generated using tools from the FTOOL package, based on a count rate threshold of 3.5–10.0 $\rm counts\,s^{-1}$ (See Figure \ref{fig9:nicer_mode_selection}). Subsequently, the task \textsc{nicerl3-spect} was used to extract the source and background spectra for the high-mode, along with the associated ancillary response file (ARF) and redistribution matrix file (RMF), using the high-mode GTI. During both temporal and spectral analysis, background contributions were estimated using the SCORPEON model, incorporating the latest available geomagnetic data to ensure accurate background correction.

%%%%%%%%%%%%%%%%%%%%%%%%%%%%%%%%%%%%%%%%%%%%%%%%%%%%%%%%%%%%%%%%%%%%%%%%%%%%%%%%%%%%%%%%%%%%%%%%%%%%%%%%%%%%%%%%%%%%%%%%%%%%%%%%%%%%%%%%%%%%%%%%%%%%%%%%%%%%%%%%%%%%%%%%%%%%%%%%%%%%%%%%%%%%%%%%%

\subsection{Chandra}
PSR J1023+0038 was observed with the Chandra X-ray Observatory on October 16 as part of a targeted Guest Observer (GO) program aimed at capturing its high-time-resolution behaviour. For this study, we selected this observation due to its use of Continuous Clocking (CC) mode and sufficient exposure time, which together provide the necessary temporal resolution and photon statistics for detailed spectral and timing analysis.
\subsubsection{Chandra observation and data reduction}
PSR J1023+0038 was observed by Chandra for a duration of 20 ks (ObsId 17785) using the back-illuminated Advanced CCD Imaging Spectrometer \citep[ACIS;][]{Garmire2003} in Continuous Clocking (CC33\_GRADED) mode, which provides a high temporal resolution of 2.85 ms by sacrificing imaging in one spatial dimension with FOV of $\approx 8.3^{'} \times 50.6^{'}$. This configuration is particularly suited for studying rapid variability in bright X-ray sources while minimising photon pile-up effects. The observation was carried out using the ACIS-S3 chip, having $1024\times1024$ pixels with a pixel size of approximately 0.492 arcseconds, providing high spatial resolution with a point spread function of $\sim$1 arcsecond.

Data reduction of PSR J1023+0038 was performed with the standard Chandra Interactive Analysis of Observations \textsc{CIAO version 4.17.0} using the Calibration Database \textsc{CALDB version 4.12.0}, following the standard data reduction science threads. The processing of Level=1 observation files was carried out using the CIAO task \textsc{chandra\_repro}  to generate new Level-2 event files with the latest calibration applied. Due to the CC-mode setup, spatial filtering was limited to a one-dimensional strip along the readout direction. Source events were extracted from a square region of $6^{''}\times6^{''}$, centred on the source position, while background events were selected from a nearby, source-free region of the same size. 
To identify high-mode intervals, a light curve was generated in the 0.5–7.5 keV energy range with 30-s time binning using the CIAO tool \textsc{dmextract}. Good Time Intervals (GTIs) corresponding to the high-mode were then created by applying a count rate filter in the range of 0.6–1.7 $\rm counts\,s^{-1}$ using the task \textsc{dmgti} (See Figure \ref{fig12:chandra_mode_selection}). Subsequently, the filtered event list was used to extract the high-mode source and background spectra with the \textsc{specextract} task, which also generated the corresponding Auxiliary Response File (ARF) and Redistribution Matrix File (RMF). All extracted products were grouped to a minimum of 20 counts per energy bin for spectral analysis.

%%%%%%%%%%%%%%%%%%%%%%%%%%%%%%%%%%%%%%%%%%%%%%%%%%%%%%%%%%%%%%%%%%%%%%%%%%%%%%%%%%%%%%%%%%%%%%%%%%%%%%%%%%%%%%%%%%%%%%%%%%%%%%%%%%%%%%%%%%%%%%%%%%%%%%%%%%%%%%%%%%%%%%%%%%%%%%%%%%%%%%%%%%%%%%%%%%%%%%
\section{Analysis and results}\label{Results}

\subsection{\xmm{}}
\subsubsection{X-ray variability and Mode selection of \xmm{}}
\label{xmm_mode_selection}
%%%%%%%%%%%%%%%%%%%%%%%%%%%%%%%%%%%%%%%%%%%%%%%%%%%%%%%%%%%%%%%%%%%%%%%%%%%%%%%%%%%%%%%%%%%%%%%%%
\begin{figure*}[ht]
    \centering
    % First row
    \includegraphics[width=0.49\textwidth]{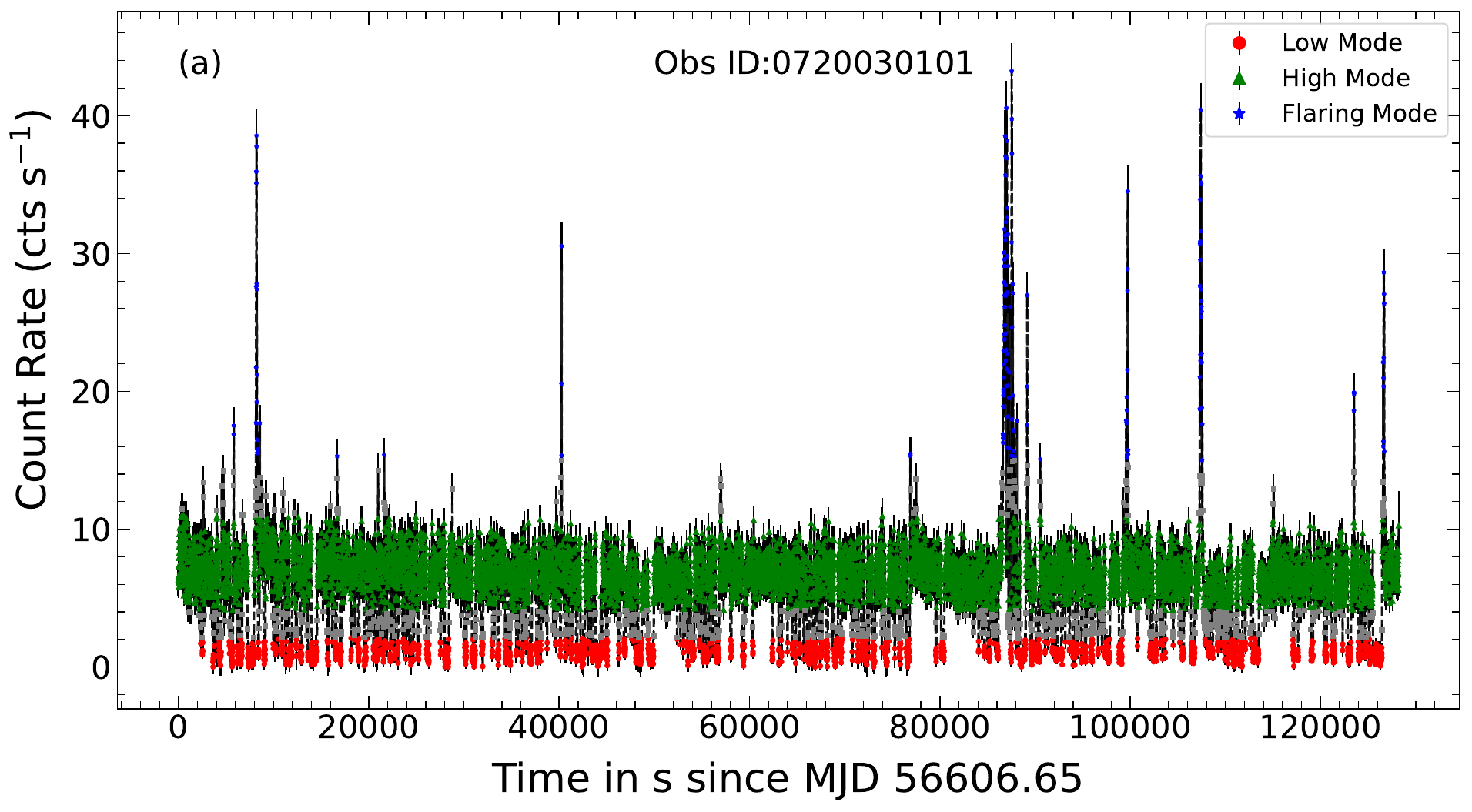}
    %\hfill
    \includegraphics[width=0.49\textwidth]{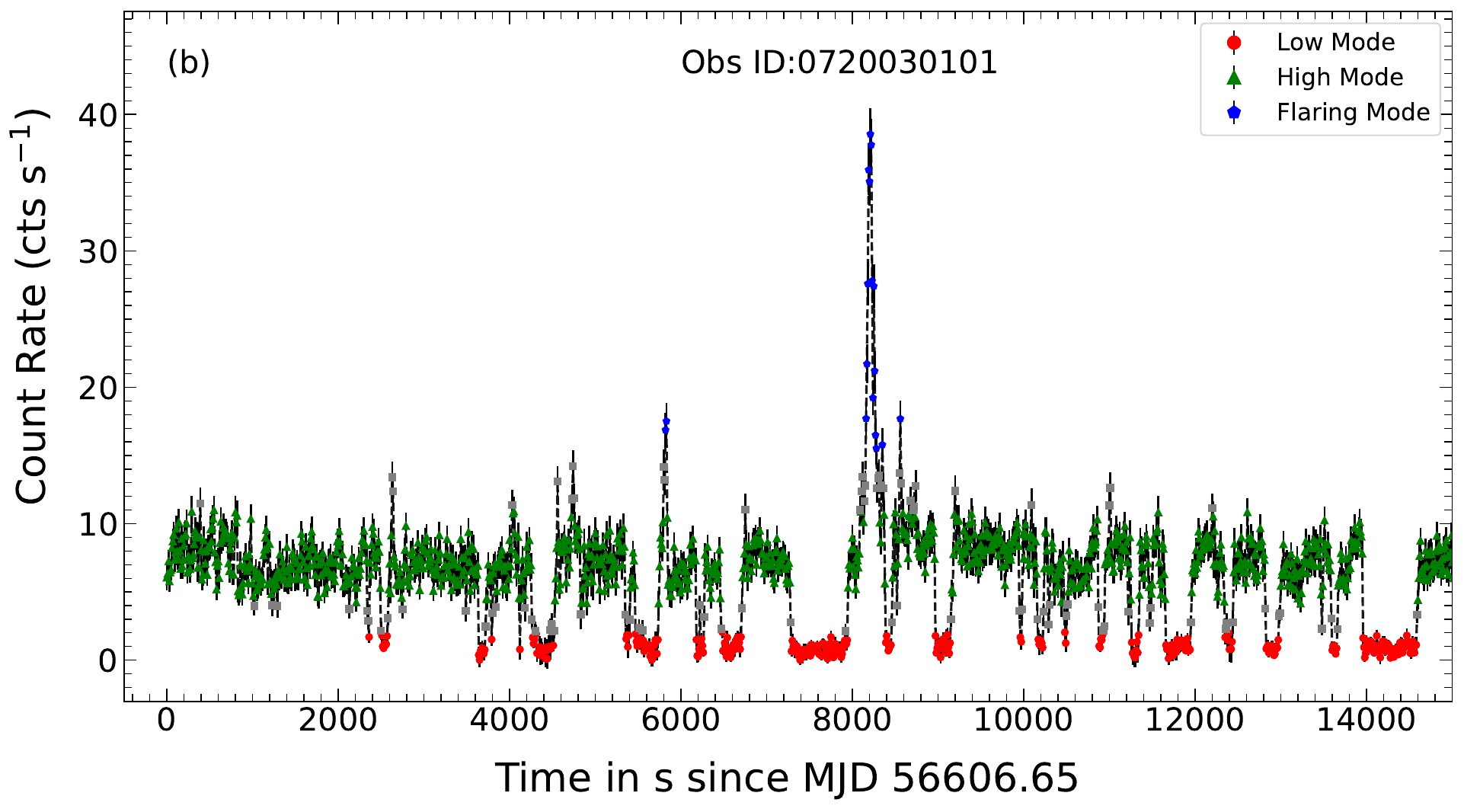}
    
    %\vspace{0.5cm}
    
    % Second row
    \includegraphics[width=0.50\textwidth]{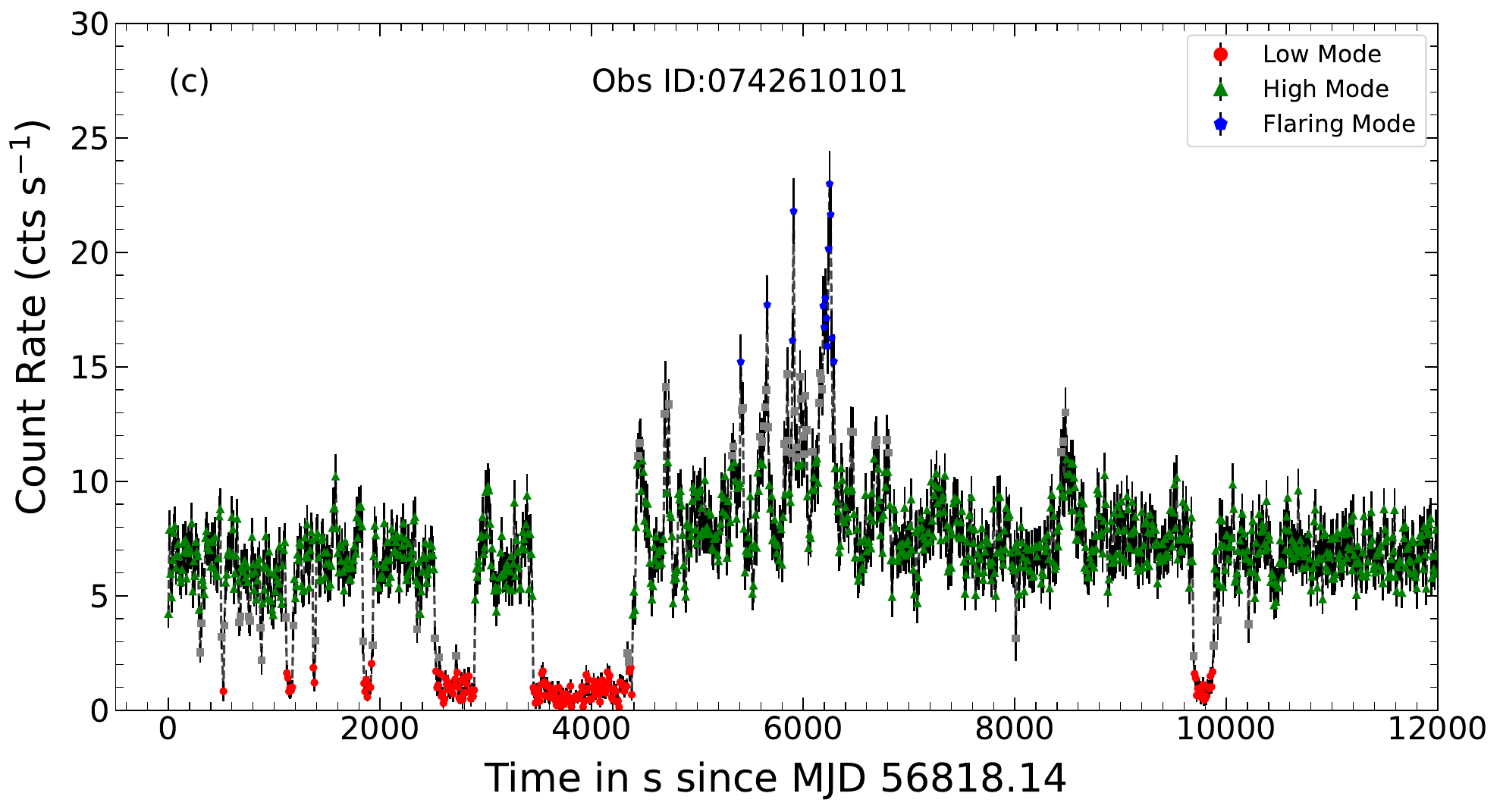}
    %\hfill
    \includegraphics[width=0.49\textwidth]{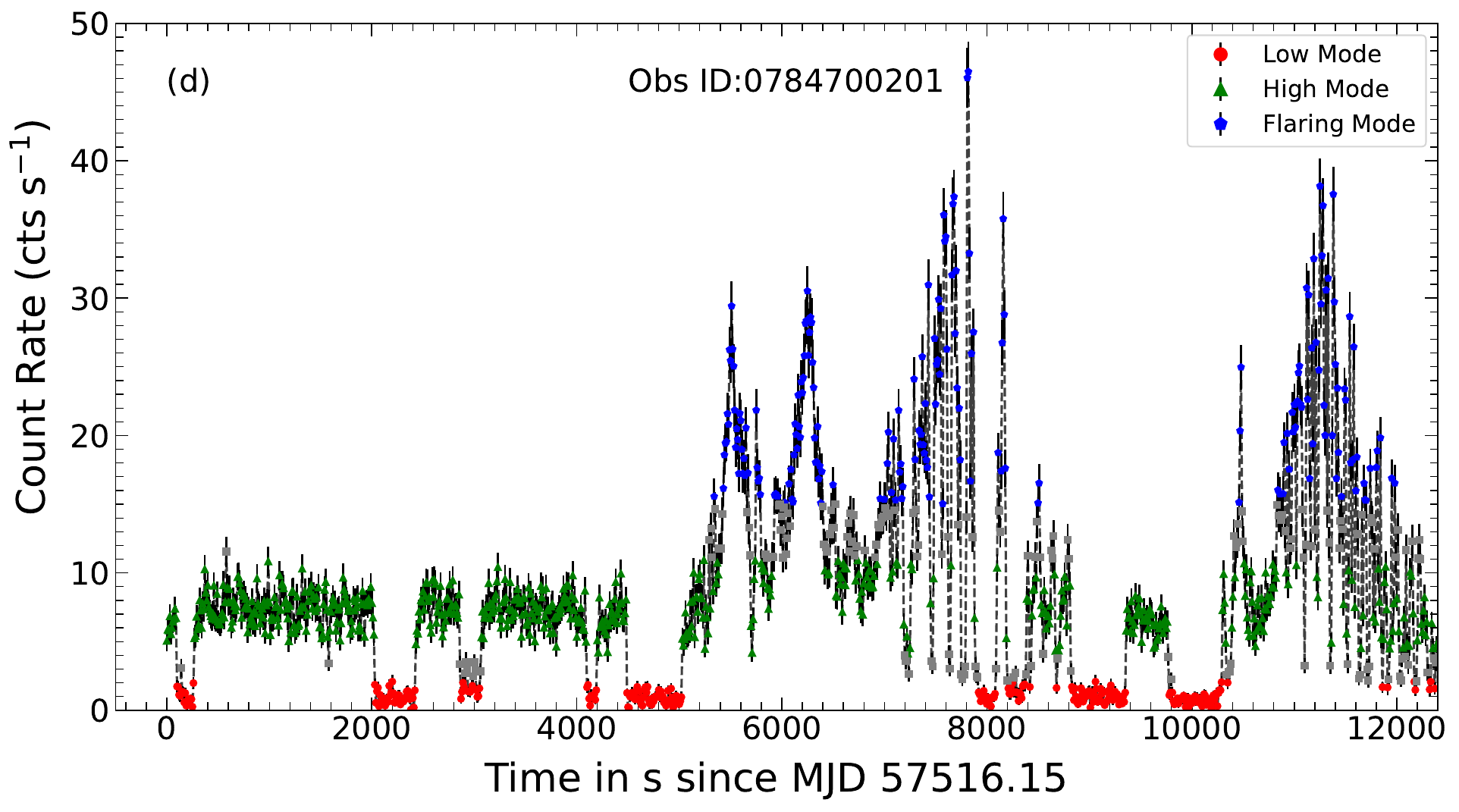}
    
    \caption{\xmm{} EPIC co-added, background-subtracted, and exposure corrected X-ray light curves in 0.3--10.0 keV energy range binned at a time resolution of 10s. (a) total light curve from ObsID 0720030101, (b) Zoom-in light curve from ObsID 0720030101, (c) Zoom-in light curve from ObsID 0742610101, (d) Zoom-in light curve from ObsID 0784700201.}
    \label{fig1:xmm_long_lc}
\end{figure*}
%%%%%%%%%%%%%%%%%%%%%%%%%%%%%%%%%%%%%%%%%%%%%%%%%%%%%%%%%%%%%%%%%%%%%%%%%%%%%%%%%%%%%%%%%%%%%
Figure \ref{fig1:xmm_long_lc} shows the \xmm{} EPIC background-subtracted and exposure-corrected X-ray light curves (Total and zoom-in) in the 0.3–10.0 keV energy band, obtained by combining data from the MOS1, MOS2, and PN instruments across the three most prolonged observations, with a time binning of 10 s.

\begin{figure*}[ht]
    \centering
    % First row
    \hspace{1.05cm}
    \raisebox{-6.5cm}{\includegraphics[width=0.43\textwidth, height=0.23\textheight]{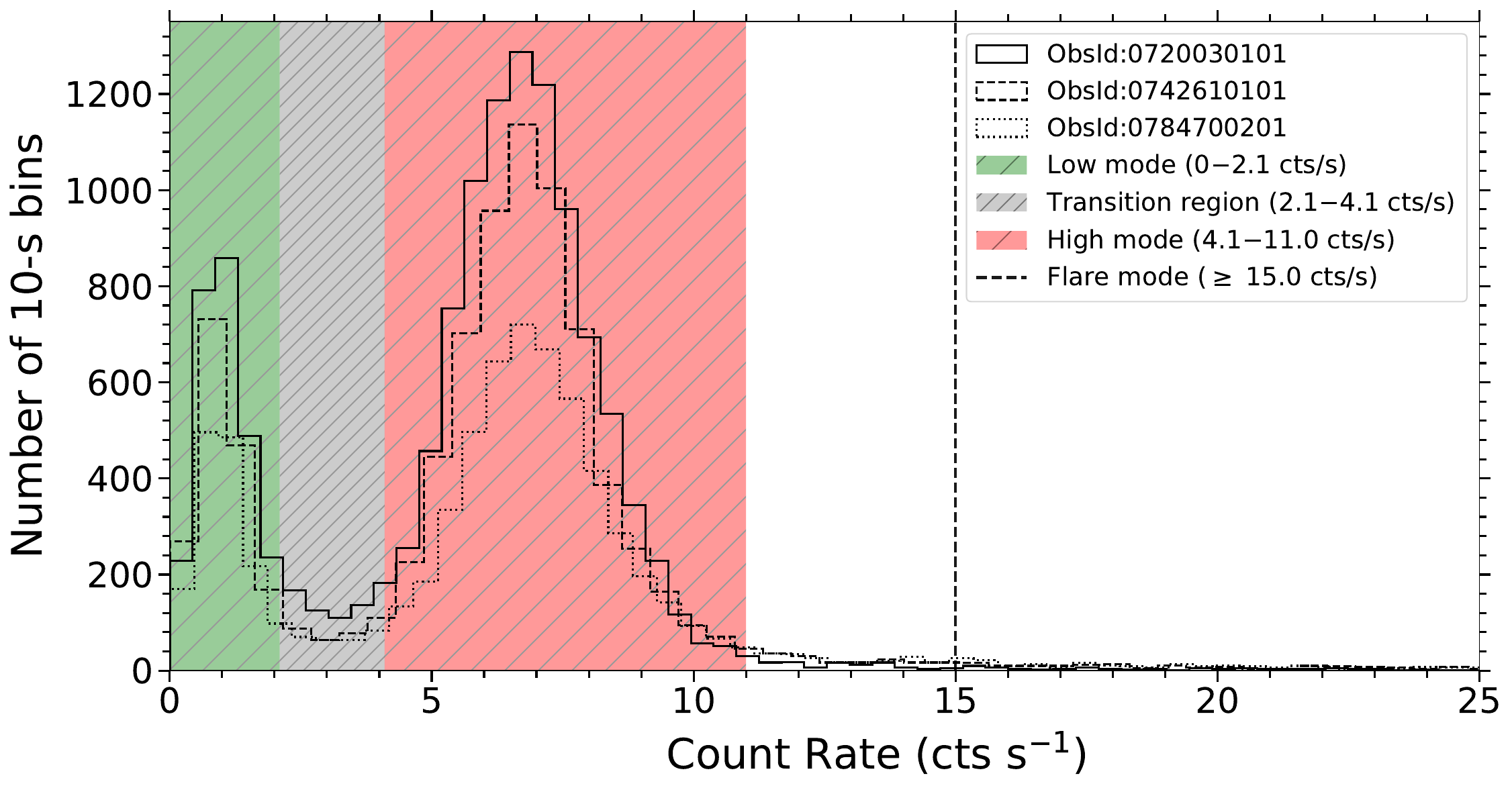}}
    \hfill
    \includegraphics[width=0.35\textwidth,angle=270]{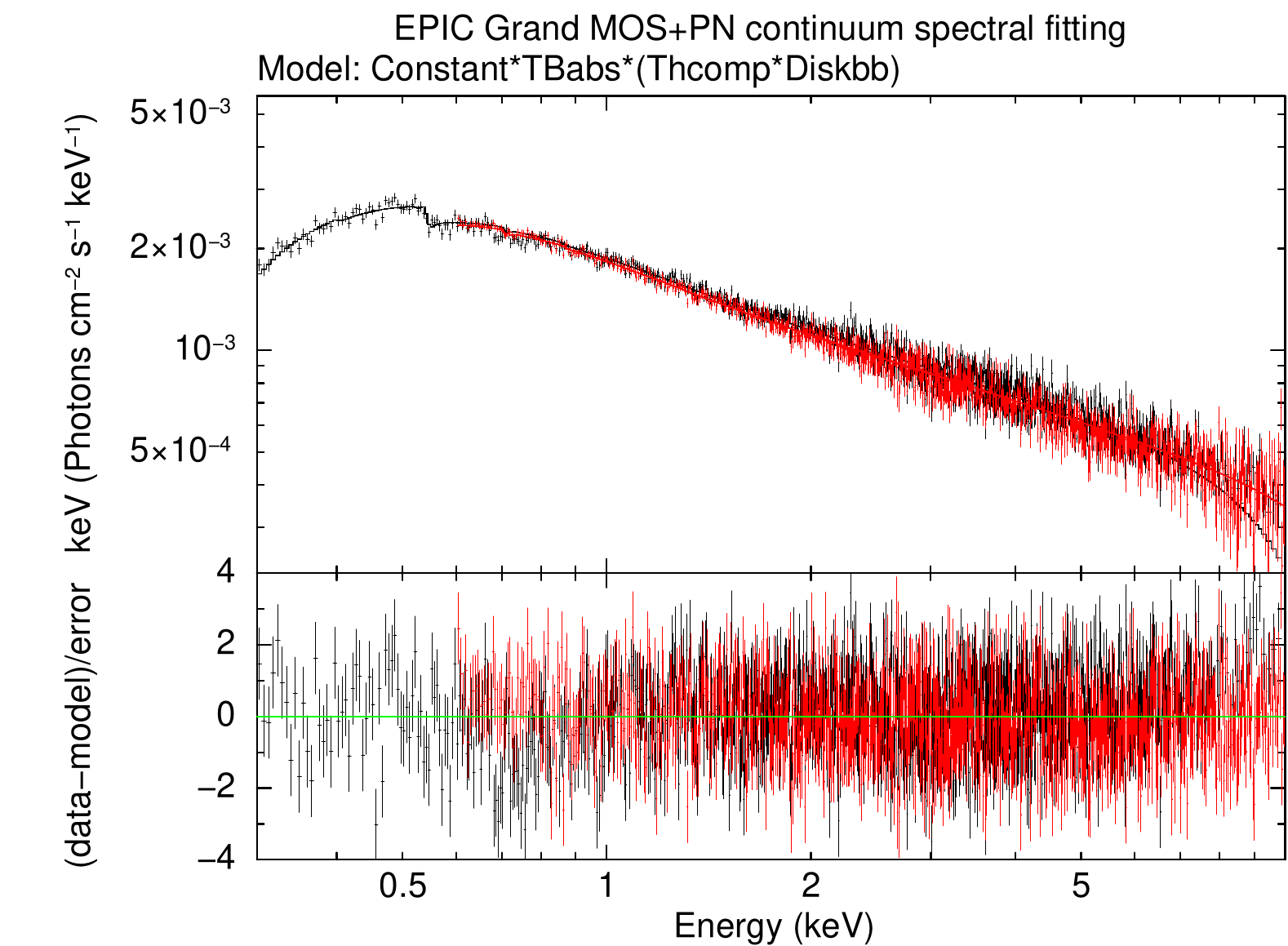}
    
    % Second row
    \hspace*{0.03\textwidth}
    \includegraphics[width=0.37\textwidth,angle=270]{Fig3b-xmm_T_gamma_review.eps}
    \hfill
    \includegraphics[width=0.37\textwidth,angle=270]{Fig3c-xmm_T_norm_review.eps}
    
     \caption{The top left panel shows the distribution of count rates produced from the 10-s binned light curves from the three longest observations of \xmm{} taken between November 2013 and May 2016. The top right panel shows the unfolded \xmm{} EPIC Grand MOS and PN spectra in the 0.3–10.0 keV (MOS) and 0.6-10.0 keV (PN) energy range, respectively, along with the residuals. The spectra are best fitted with an absorption model (TBabs), a disc blackbody model (diskbb), and a thermally Comptonized continuum model (Thcomp). The bottom left panel displays the integrated confidence contours between the power-law photon index ($\Gamma_{\tau}$) and disc temperature ($kT_{in}$), obtained from the EPIC Grand MOS+PN spectral analysis. The bottom right panel shows similar confidence contours for the power-law photon index ($\Gamma_{\tau}$) and disc normalisation. The contours represent 90\% (red), 95.5\% (green), and 99\% (blue) confidence levels. }
    \label{fig2:xmm_long_spec}
\end{figure*}

All these \xmm{} observations of PSR J1023+0038 exhibit rapid X-ray variability and display three distinct emission modes: high, low, and flare. Among these, the high-mode is the most persistent, with an average count rate of $\rm \approx 7 \,counts\,s^{-1}$. Transitions to the low-mode occur unpredictably, with the count rate dropping to about $\rm \approx 1 \, counts\,s^{-1}$ within 10–30 s. Intense X-ray flares are observed roughly every few hours, with peak count rates reaching up to $\rm \approx 60 \,counts\,s^{-1}$, consistent with the previous studies \citep{Bogdanov2015,Archibald2015}. For the mode selection of \xmm{} observations, we followed the same approach and analysis proposed by \citet{Bogdanov2015}. The top left panel of Figure \ref{fig2:xmm_long_spec} shows a clear bimodal distribution of the count rates with distinct low and high-modes across all three long exposure observations. Therefore, we define the high-mode in 4.1–11 counts s$^{-1}$, the low-mode in 0–2.1 counts s$^{-1}$, and the flaring-mode in $\geq$15 counts s$^{-1}$, similar to \citet{Bogdanov2015}.

%%%%%%%%%%%%%%%%%%%%%%%%%%%%%%%%%%%%%%%%%%%%%%%%%

\subsubsection{Spectral analysis and results of \xmm{}}
\label{xmm_spectral_analysis}

The spectral datasets obtained from the three longest \xmm{} observations of PSR J1023+0038 served as an initial testbed for investigating the theory of partial penetration of the accretion disc into the pulsar's magnetosphere to explain the phenomenon of coherent X-ray pulsation and X-ray emission during the high-mode \citep[See][]{Campana2016,Bozzo2018,Bhattacharyya2020}. These observations were subsequently combined to enhance the statistical significance and robustness of the results obtained from the spectral analysis (See Section \ref{xmm_observation & data reduction}). To derive the spectral parameters of PSR J1023+0038, simultaneous spectral fitting was performed on the EPIC grand MOS and PN spectra in the 0.3–10.0 keV and 0.6–10.0 keV energy bands, respectively. The fitting employed a composite model implemented in the X-ray spectral fitting package XSPEC version 12.14.1 \citep{Arnaud1996}. 

As described above, we modelled the high-mode spectra of PSR J1023+0038 using a thermally comptonised continuum convolution model \citep[Thcomp;][]{Zdziarski2020}, instead of a phenomenological power-law model. This approach provides a physically motivated description of the compotonised high-energy emission attributed to the presence of a hot corona above the accretion disc or to shocks produced by the propeller mechanism acting on the infalling material. To account for the soft X-ray emission and upscattered seed photons, we included a thermalised multi-colour accretion disc \citep[Diskbb;] []{Mitsuda1984,Makishima1986} component representing a geometrically thin, optically thick accretion disc. Additionally, an X-ray absorption cross-section, modelled as the Tuebingen-Boulder ISM absorption \citep[TBabs;][]{Wilms2000} was applied to correct the interstellar and local absorption effects along the line-of-sight by utilizing the cross section from \citet{Verner1996}, and abundances from \citep{Wilms2000}. A constant model was also implemented for the cross-calibration of MOS and PN spectra. Taking this into account, we fitted a model defined as \textsc{Constant*TBabs*(Thcomp$\otimes$Diskbb)} and obtained the best-fit model parameters for the observed X-ray spectra using a chi-square minimisation algorithm. The top right panel of Figure \ref{fig2:xmm_long_spec} shows the best-fitted model along with the residual, and Table \ref{tab:spectral_param} summarises the best-fit spectral parameters obtained from the spectral analysis. The high-mode spectral data give a relatively good fit for the defined model with $\rm \chi^2_{red} = 1.06$ for 2057 degrees of freedom with a null hypothesis probability of 2.6\%. To assess the statistical robustness of the proposed model, confidence contours at significance levels of 90\% ($1.6 \sigma$), 95.5\% ($2.0 \sigma$) and 99\% ($2.6 \sigma$) were generated between key model parameters: the disc temperature, disc normalisation, and power-law photon index using the \textsc{steppar} task within XSPEC package. These contours are shown in the bottom panels of Figure \ref{fig2:xmm_long_spec}.

The spectral analysis indicates the presence of a high-energy Comptonised corona surrounding the accretion disc, characterised by a power-law photon index of $\Gamma_{\tau} = 1.66^{+0.01}_{-0.01}$, and up-scatters the soft seed photons originating from the accretion disc to higher energies. In addition, the soft thermal component from the accretion disc, with a temperature of $0.21^{+0.01}_{-0.01}$ keV, is well supported by the data. To statistically justify the inclusion of this component, the information criteria such as Akaike Information Criterion \citep[AIC;][]{Akaike1974} and Bayesian Information Criterion \citep[BIC;][]{Schwarz1978} were used to compare our model (i.e., Model A) with \textsc{Constant*TBabs*(Thcomp$\otimes$bbodyrad)} (i.e., Model B) \citep{Liddle2007}, which represents a single-temperature blackbody emission component (bbodyrad) as the source of seed photons. The model comparison based on the AIC and BIC is presented in Table \ref{tab:disk-significance}, showing a strong preference for our model with $\Delta \mathrm{AIC_{AB}}$ and $\Delta \mathrm{BIC_{AB}} \gtrsim 115$ \citep{Shi2012}, indicating that the \textsc{diskbb} model provides a better statistical description of the data.

Considering the source distance (D) of $1.37_{-0.04}^{+0.04}$ kpc \citep{Deller2012}, an inclination angle ({\it i}) of $42_{-2}^{+2}$ degrees \citep{Archibald2009,Archibald2013}, colour-correction factor ($\kappa$) of $1.85_{-0.15}^{+0.15}$ [i.e.,1.7--2.0] \citep{Shimura1995} and general relativity correction factor $(\xi)$ of $0.412$ \citep{Kubota1998}, we calculated the apparent inner disc radius of PSR J1023+0038 from the observed data during the high-mode using the following relations: 
\begin{equation}\label{equation1}
r_{\rm in} = \sqrt{\frac{N_{\rm disc}}{\cos i}} \times \frac{D}{10~\mathrm{kpc}},
\end{equation}
where $N_{\rm disc}$ is the normalisation of the \textsc{diskbb} model component.The true inner disc radius 
($\rm R_{\mathrm{in}}^{\mathrm{true}}$) is defined as 
\begin{equation}\label{equation2}
\rm R^{true}_{\text{in}} = \xi*\kappa^2*r^{'}_{in}
\end{equation}
where $r^{'}_{\rm in}$ is the corrected apparent inner disc radius, given as 
\begin{equation}\label{equation3}
\rm r^{'}_{\rm in} = p*r_{in}
\end{equation}

The correction factor $\rm p$ was determined based on the empirical relation between the compact object mass ($\rm M_*$) and the apparent inner disc radius ($\rm r_{\rm in}$) provided by \citet{Kubota1998}, expressed as $\rm M_*/r_{in} \sim 0.13-0.58 \,M_\odot\,km^{-1}$ with lower the compact object masses, smaller values of $\rm M_*/r_{in}$ are preferred \citep{Yaqoob1993}. Following this relation, for a canonical neutron star mass of 1.4 $\rm M_\odot$, the apparent disc radius is found to be 10.76 km \citep[][and references therein]{Kubota1998}.
For PSR J1023+0038, with $\rm M_* = 1.71 \pm 0.16 \,M_\odot$ \citep{Archibald2009, Deller2012}, the estimated apparent radius ($\rm r_{in}^E$) is $13.15^{+1.2}_{-1.2}$ km. In contrast, the apparent disc radii derived from the observed normalization, inclination, and distance ($\rm r_{\rm in}$), using Eq. \ref{equation1}, lie within the range 1.3--3.0 km. Similar discrepancies have also been reported in \citet{Campana2016} and \citet{CotiZelati2018}. They have used a correction factor of 2-3, including the $\xi$ factor. Without $\xi$ such a correction factor will be around 4.8--7.3 for 1.4 $M_\odot$.
Using the compact object mass of 1.71 $\pm$ 0.16 $M_\odot$, we have estimated and used a similar correction factor, p, to calculate the true radius R$^{true}_{in}$. Considering all observations from different instruments in this work, the mean value of $\rm p$ is estimated to be 7.0. The true inner disc radius during the high-mode ($\rm R_{\mathrm{in}}^{\mathrm{true}}$) was estimated using the equation \ref{equation2} and summarised in Table \ref{tab:disk-significance}.

The estimated inner disc radius from \xmm{} is $16.8_{-3.8}^{+3.8}$ km (3$\sigma$ significance). This finding supports the scenario in which the accretion disc extends inside the magnetosphere, as mentioned in  \citet{Campana2016,Bozzo2018,Bhattacharyya2020}.
%%%%%%%%%%%%%%%%%%%%%%%%%%%%%%%%%%%%%%%%%%%%%%%%%%%%%%%%%%%%%%%%%%%%%%%%%%%%%%%%%%%%%%%%%%%%%%%%%%%%%%%%%%%%%%%%%%%%%%%%%%%%%%%%%%%%%%%%%%%%%%5
\begin{figure*}[ht]
    \centering
    % First row
    \includegraphics[width=0.48\textwidth]{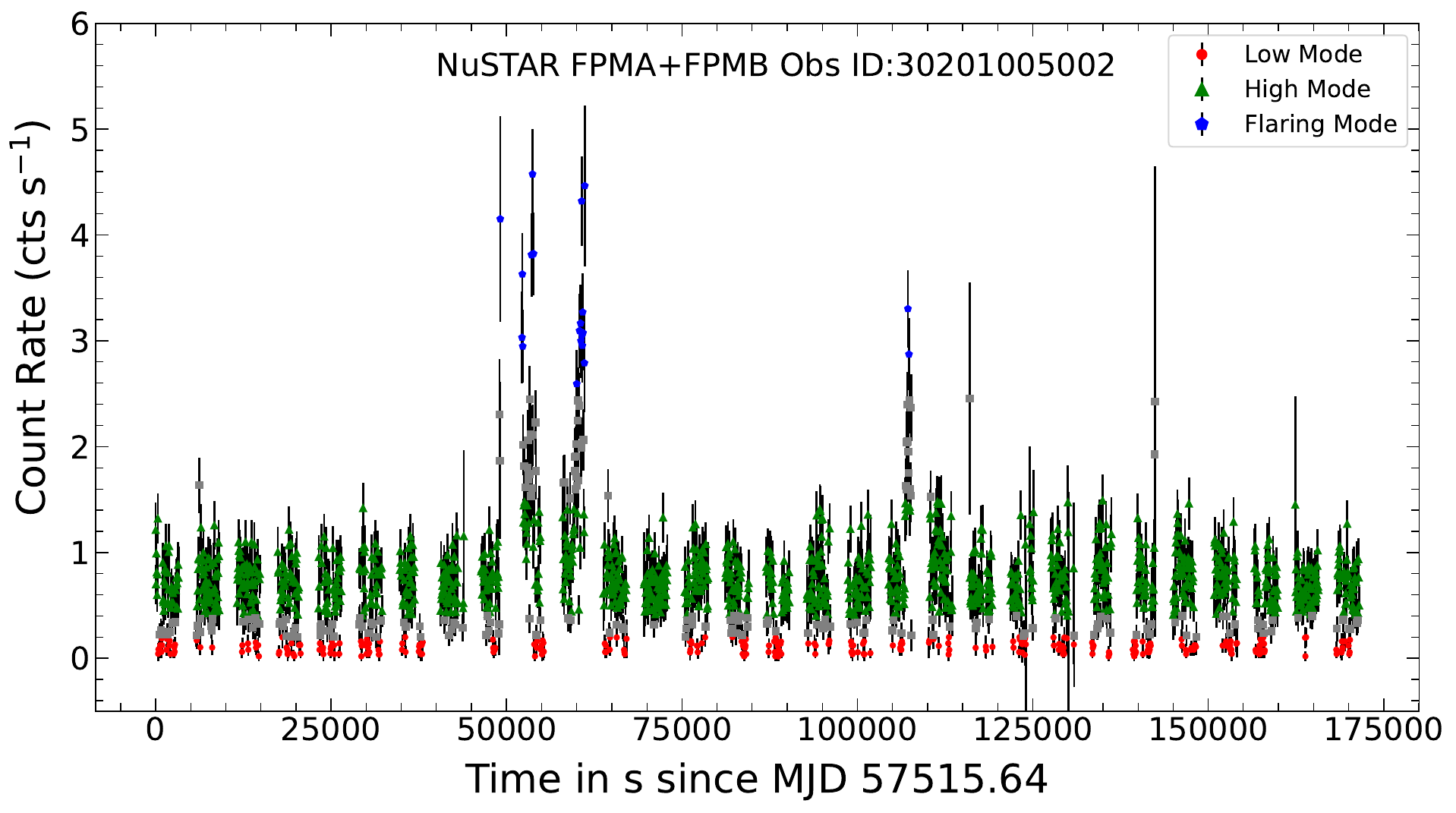}
    \includegraphics[width=0.51\textwidth]{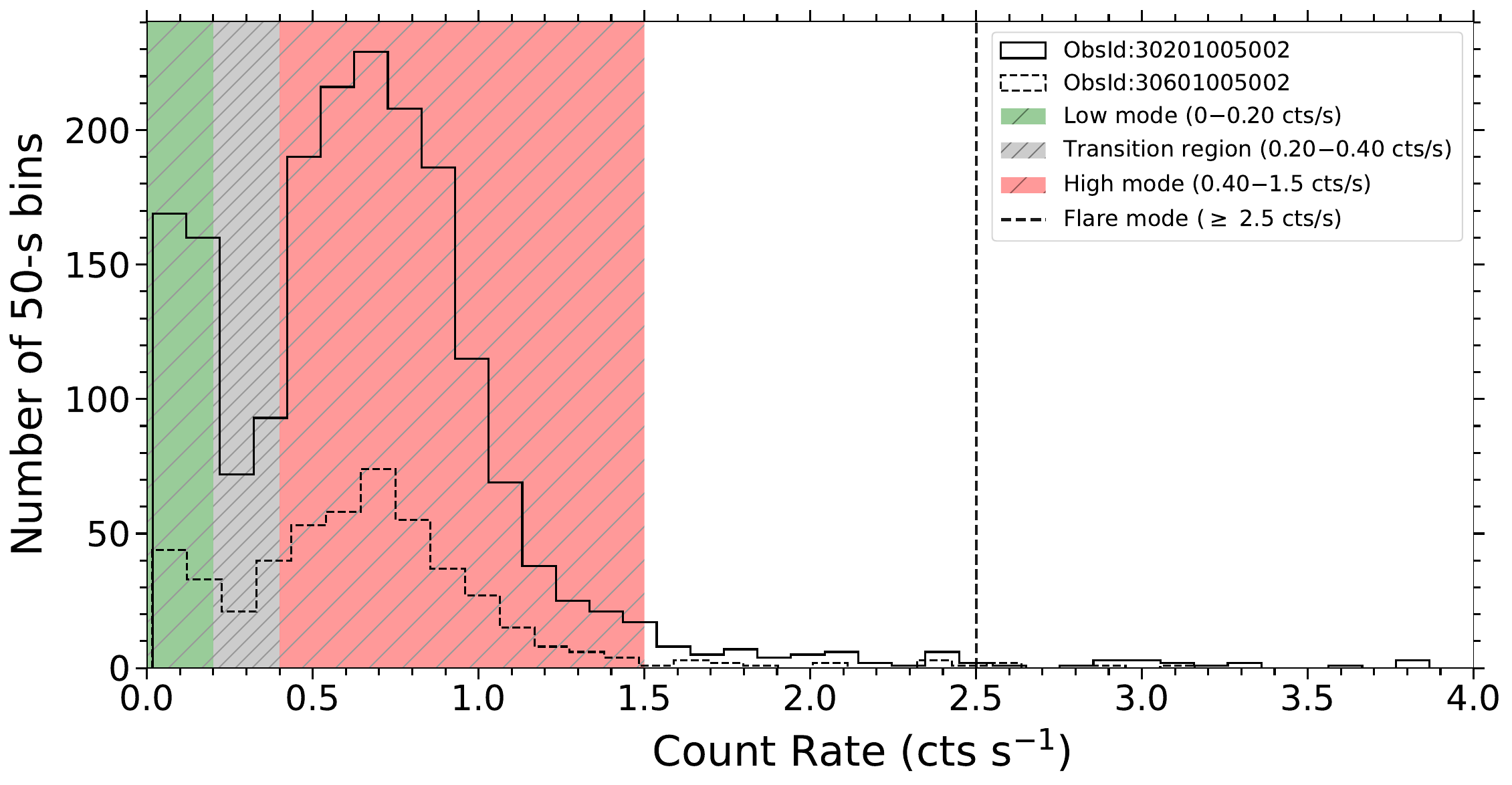}
    \caption{Left: \nus{} (FPMA+FPMB) co-added and background-subtracted X-ray light curve of ObsId:30201005002 in 3.0–79.0 keV energy range binned at a time resolution of 50s. Right: Distribution of count rates produced from the 50-s binned light curves from the two simultaneous observations of \nus{} taken in May 2016 and June 2021.}
    \label{fig4:nustar_mode_selection}
\end{figure*}
%%%%%%%%%%%%%%%%%%%%%%%%%%%%%%%%%%%%%%%%%%%%%%%%%%%%%%%%%%%%%%%%%%%%%%%%%%%%%%%%%%%%%%%%%%
\begin{figure}[ht]
    \begin{flushleft}
        \centering
        \includegraphics[width=0.35\textwidth,angle=270]{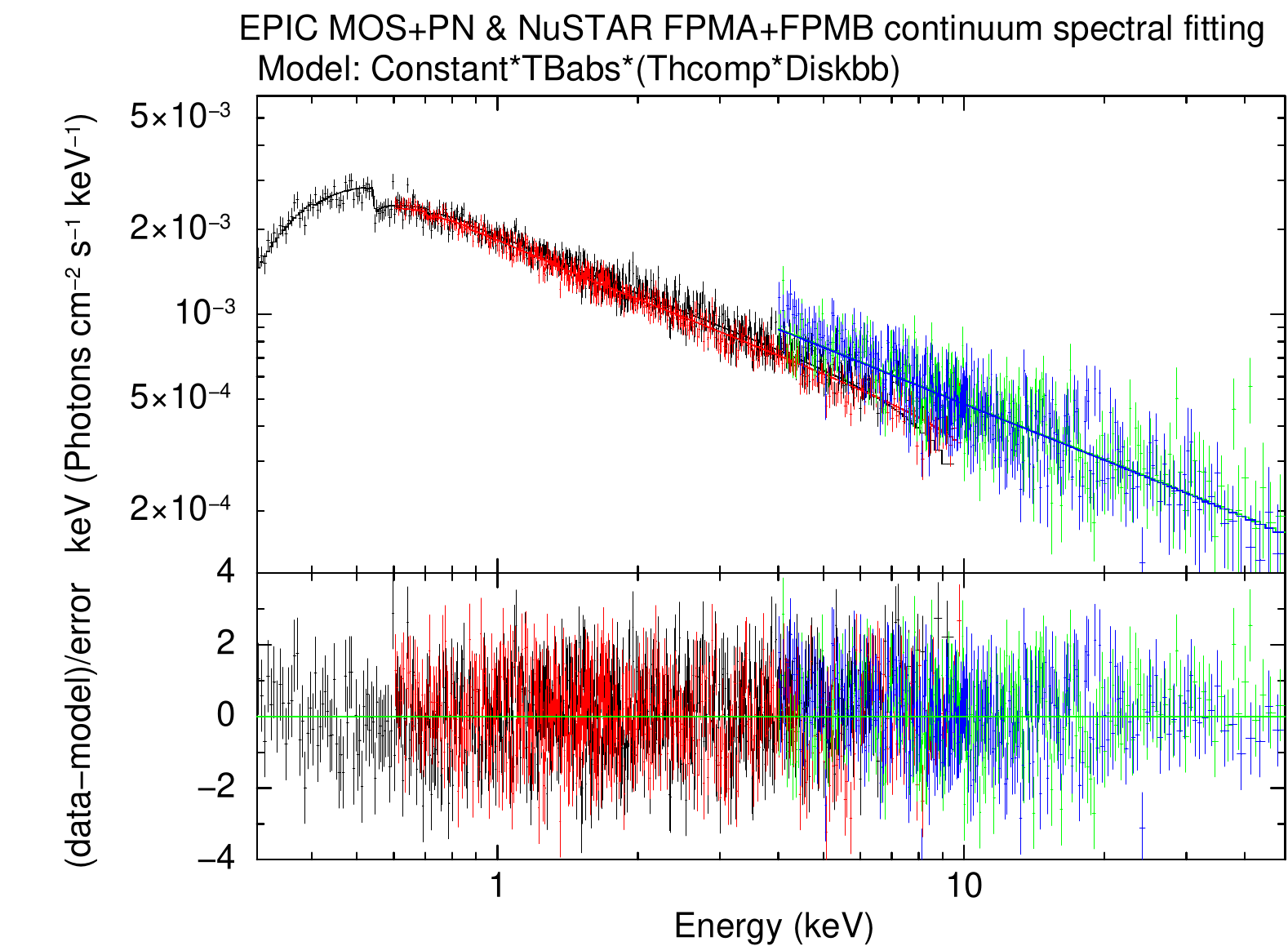}
        \caption{Top panel shows joint \xmm{} and \nus{} spectral fit for PSR J1023+0038 during the semi-simultaneous (N1+X3) observation. The bottom panel shows the residuals of the fitting.}
        \label{fig5:epic_nustar_semi_sim_spec}
    \end{flushleft}
\end{figure}
\begin{figure}[ht]
    \begin{flushright}
        \centering
        \includegraphics[width=0.35\textwidth,angle=270]{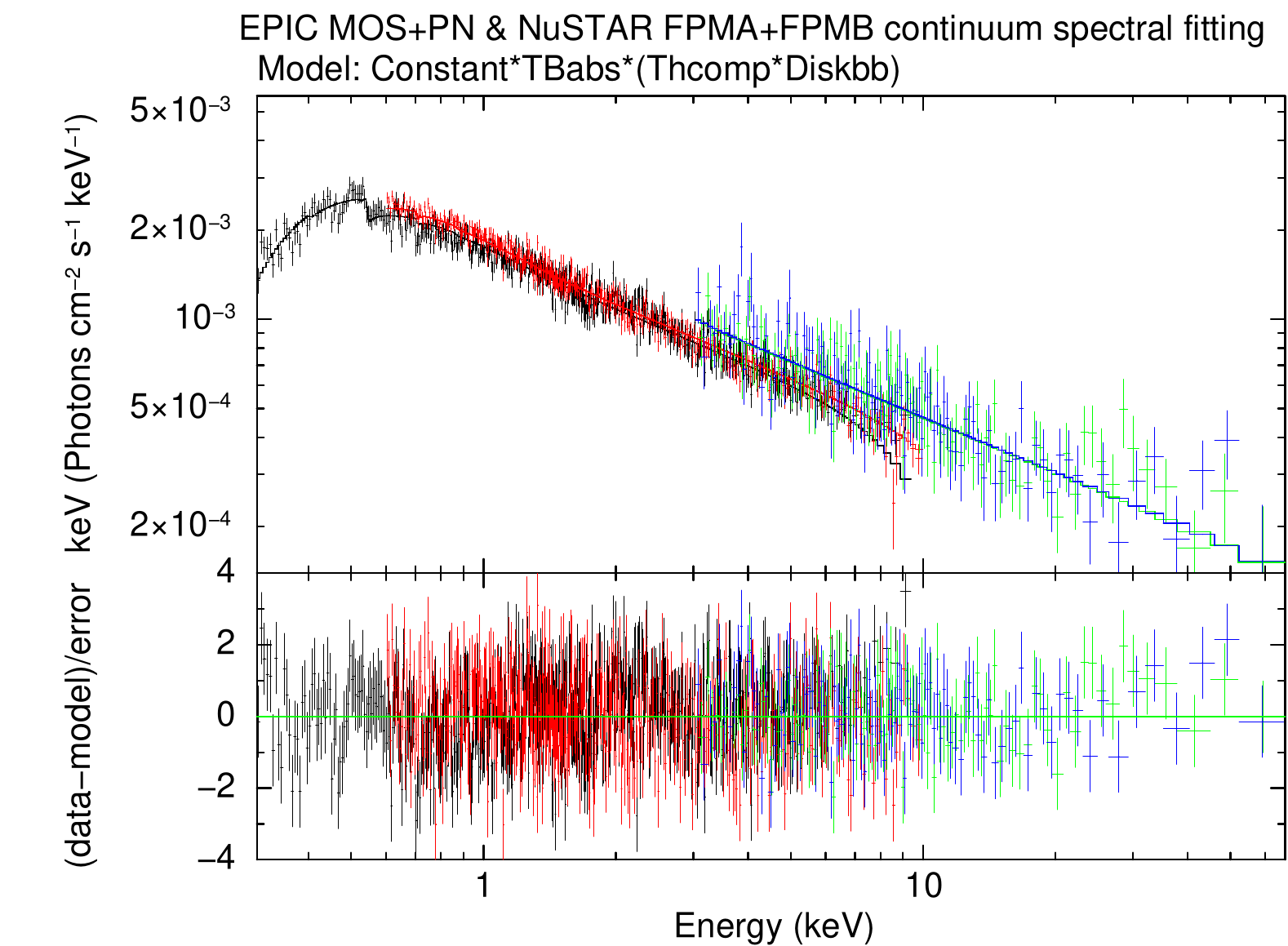}
        \caption{Top panel shows joint \xmm{} and \nus{} spectral fit for PSR J1023+0038 during the simultaneous (N2+X4) observation. The bottom panel shows the residuals of the fitting.}
        \label{fig6:epic_nustar_sim_spec}
    \end{flushright}
\end{figure}
%%%%%%%%%%%%%%%%%%%%%%%%%%%%%%%%%%%%%%%%%%%%%%%%%%%%%%%%%%%%%%%%%%%%%%%%%%%%%%%%%%%%

%%%%%%%%%%%%%%%%%%%%%%%%%%%%%%%%%%%%%%%%%%%%%%%%%%%%%%%%%%%%%%%%%%%%%%%
\begin{figure*}[ht]
    \centering
    \includegraphics[width=0.37\textwidth,angle=270]{Fig7a-xmm_nustar_N1_X3_gamma_Tin_review.eps}
    \includegraphics[width=0.37\textwidth,angle=270]{Fig7b-xmm_nustar_N1_X3_gamma_tnorm_review.eps}
     \caption{The contours represent 90\% (red), 95.5\% (green), and 99\% (blue) confidence levels of best-fit spectral parameters for \nus{} and XMM Newton semi-simultaneous (N1+X3) observations. The left shows integrated confidence contours between the power-law photon index ($\Gamma_{\tau}$) and disc temperature ($kT_{in}$), and the right panel shows similar confidence contours between the power-law photon index ($\Gamma_{\tau}$) and disc normalisation.  }
    \label{fig7:xmm_nustar_semi_simultaneous_contours}
\end{figure*}
\begin{figure*}[!ht]
    \centering
    \includegraphics[width=0.37\textwidth,angle=270]{Fig8a-xmm-nustar_xmm_N2_X4_gamma_Tin_review.eps}
    \includegraphics[width=0.37\textwidth,angle=270]{Fig8b-xmm_nustar_N2_X4_gamma_tnorm_review.eps}
     \caption{The contours represent 90\% (red), 95.5\% (green), and 99\% (blue) confidence levels of best-fit spectral parameters for \nus{} and XMM Newton simultaneous (N2+X4) observations. The left shows integrated confidence contours between the power-law photon index ($\Gamma_{\tau}$) and disc temperature ($kT_{in}$), and the right panel shows similar confidence contours between the power-law photon index ($\Gamma_{\tau}$) and disc normalisation. }
    \label{fig8:xmm_nustar_simultaneous_contours}
\end{figure*}
%%%%%%%%%%%%%%%%%%%%%%%%%%%%%%%%%%%%%%%%%%%%%%%%%%%%%%%%%%
%%%%%%%%%%%%%%%%%%%%%%%%%%%%%%%%%%%%%%%%%%%%%%%%%%%%%%%%%%%%%
\begin{figure*}[ht]
    \centering
    % First row
    \includegraphics[width=0.48\textwidth]{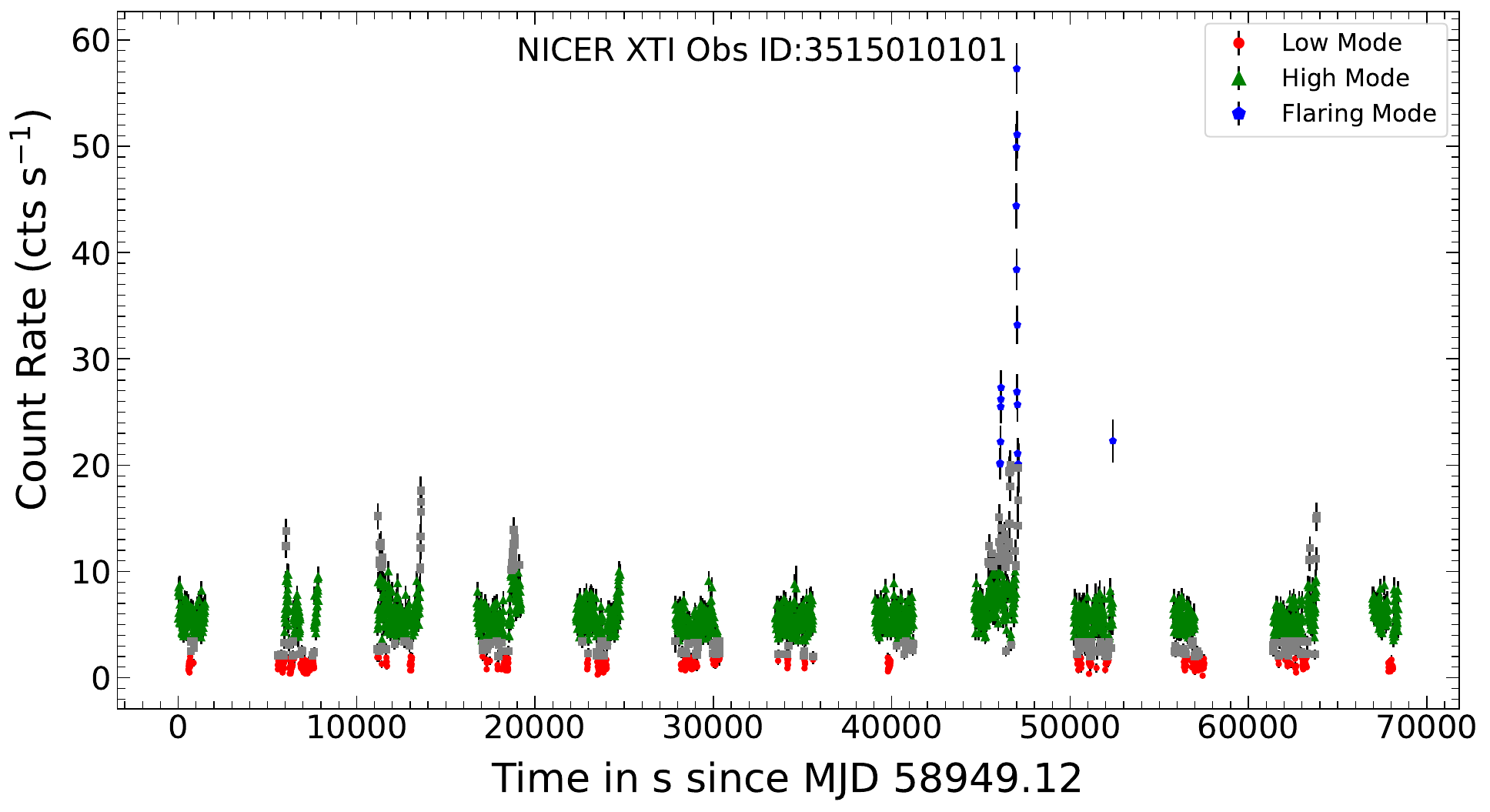}
    \includegraphics[width=0.50\textwidth]{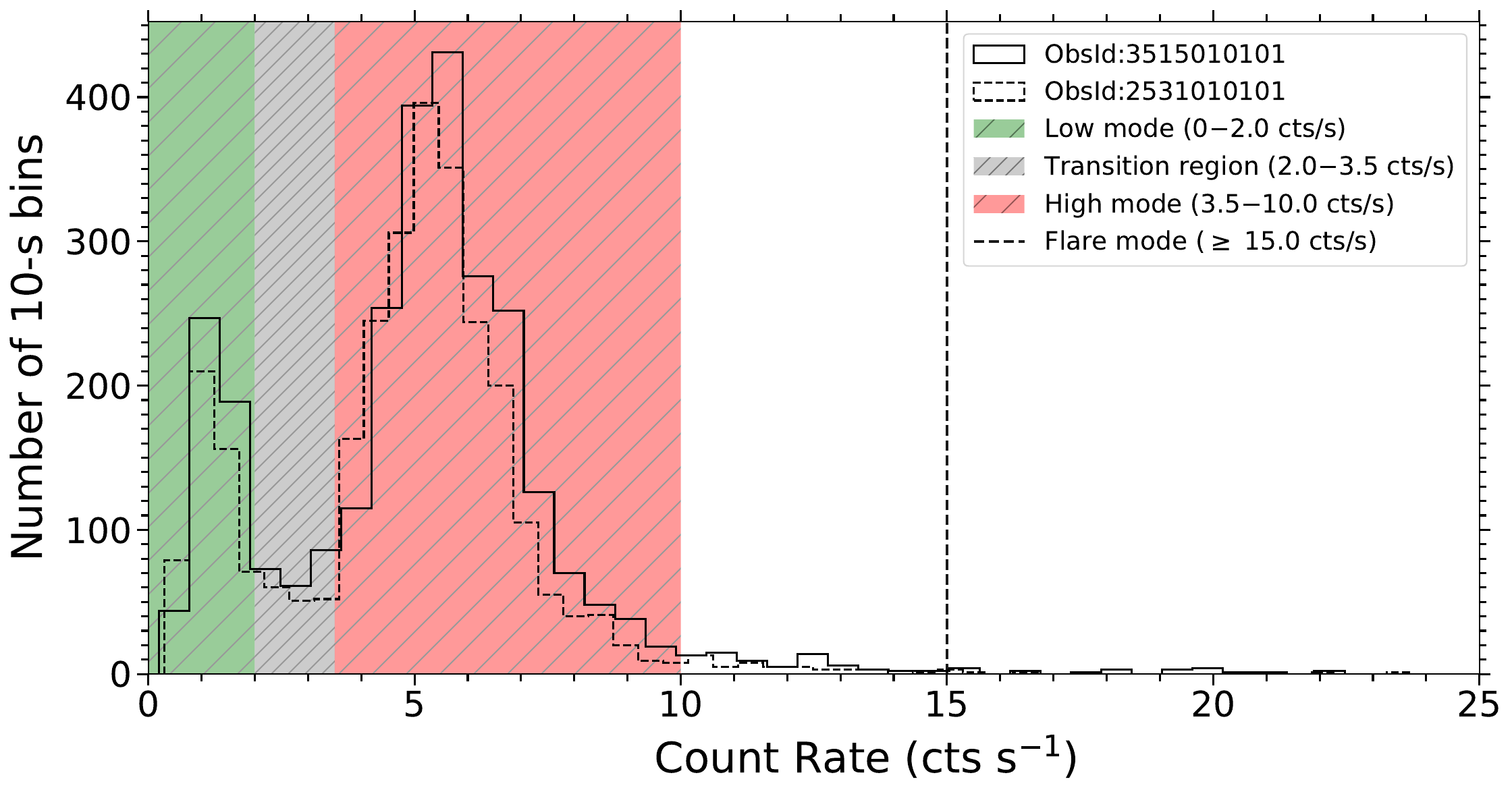}
    \caption{Left: \nicer{} XTI background-subtracted X-ray light curve of ObsId:3515010101 in 0.5--10.0 keV energy range binned at a time resolution of 10s. Right: Distribution of count rates produced from the 10-s binned light curves from the \nicer{} observation taken in April 2020 and May 2019.}
    \label{fig9:nicer_mode_selection}
\end{figure*}
%%%%%%%%%%%%%%%%%%%%%%%%%%%%%%%%%%%%%%%%%%%%%%%%%%%%%%%%%%%%%%%%%%%%
\subsection{\nus{}}
\subsubsection{X-ray variability and mode selection in \nus{}}
\label{nustar_mode_selection}
The co-added, background-subtracted light curve of PSR J1023+0038 from the \nus{}/FPMA and FPMB detectors, extracted in the 3.0–79.0 keV energy band with a time binning of 50 seconds, is presented in the left panel of Figure \ref{fig4:nustar_mode_selection}. Similar to the \xmm{} observations, both \nus{} datasets exhibit clear X-ray variability characterised by transitions between high, low, and flaring modes. In both observations, PSR J1023+0038 remains predominantly in the high-mode, with an average count rate of  $\approx\rm0.7\,counts\,s^{-1}$, and undergoes rapid transitions to the low mode within 10–30 seconds, where the average count rate drops to $\approx\rm0.1\,counts\,s^{-1}$. X-ray flaring episodes are also observed in each observation, with peak count rates reaching up to $\approx\rm5.0\,counts\,s^{-1}$.
For mode selection in the \nus{} observations of PSR J1023+0038, an analysis approach similar to that adopted by \citet{Bogdanov2015} was employed. A bimodal distribution in count rates, corresponding to the high and low modes, was prominent in both simultaneous observations, as shown in the right panel of Figure \ref{fig4:nustar_mode_selection}. Based on this distribution, we define the high-mode as intervals with count rates in the range 0.4--1.5 $\rm counts\,s^{-1}$, the low-mode as 0--0.2 counts\,s$^{-1}$, and the flaring-mode as intervals with count rates $\geq 2.5 \rm \,counts\,s^{-1}$.

\subsubsection{Joint spectral analysis and results of \nus{} and \xmm{}}
\label{nustar_spectral_analysis}
To investigate the broadband X-ray spectral properties of PSR J1023+0038 during its high-mode, we performed a joint spectral analysis using two simultaneous observations obtained with NuSTAR (ObsId: 30201005002, 30601005002) and \xmm{} (ObsId: 0784700201, 0864010101). These coordinated observations provide continuous energy coverage from 0.3 to 79.0 keV, with the EPIC MOS and PN instruments covering the soft X-ray regime and the \nus{} FPMA and FPMB instruments extending the spectral range to hard X-rays. High-mode intervals were selected from both missions using mode-resolved light curves and corresponding count rate thresholds, as described in Sections \ref{xmm_mode_selection} and \ref{nustar_mode_selection}. The MOS1 and MOS2 spectra were co-added to enhance the signal-to-noise ratio and grouped to 100 counts per energy bin, whereas the PN and \nus{} spectra were grouped to contain a minimum of 200 and 20 counts per energy bin, respectively. Spectral fitting was carried out using XSPEC version 12.14.1 \citep{Arnaud1996}, applying $\chi^2$ minimisation for parameter estimation.

To ensure consistency and physical interpretability, we adopted the same spectral model and fitting approach as described in the Section \ref{xmm_spectral_analysis}. This allowed for a direct comparison between the \xmm{}-only and the joint fits, and enabled us to assess the stability of the spectral parameters across different instruments and energy ranges. We perform the joint \xmm{}+\nus{} spectral fitting in the energy range of 0.3-50.0 keV for N1+X3 and 0.3-79.0 keV for N2+X4 observation, respectively.
The joint fit yielded statistically consistent results across both simultaneous observations. The high-energy part of the spectrum is dominated by Comptonised emission with a power-law photon index of $\Gamma_{\tau} = 1.69^{+0.01}_{-0.01}$ and $1.65^{+0.02}_{-0.01}$, while the soft thermal component is well described by an inner disc temperature ($kT_{in}$) of $0.18^{+0.01}_{-0.01}$ keV and $0.20^{+0.01}_{-0.01}$ keV, respectively.  Based on the equation \ref{equation2}, the true inner disc radius during high-mode was estimated to be $\rm R_{in}^{true}$ = $23.2^{+6.9}_{-6.6}$ and $18.6^{+6.5}_{-5.6}$ km for respective observations.

The broadband spectral model provides an excellent fit to the combined datasets, with a reduced chi-square of $\chi^2_{\nu} = 1.02$ for 1586 degrees of freedom and $\chi^2_{\nu} = 1.03$ for 1152 degrees of freedom with null hypothesis probabilities of 27.7\% and 19.8\% for both simultaneous observations, respectively. No significant residuals were observed across the entire energy range, confirming the adequacy of the model. Figure \ref{fig5:epic_nustar_semi_sim_spec} and \ref{fig6:epic_nustar_sim_spec} display the unfolded spectra with the best-fit model and corresponding residuals, and the derived best-fit spectral parameters are listed in Table \ref{tab:spectral_param}.
The consistency of spectral parameters across both observations supports the temporal stability of the high-mode emission. To validate the statistical significance of the spectral model, confidence contours were generated for the confidence level of 90\%, 95.5\%, and 99\% between the parameters such as disc temperature, disc normalisation, and power-law photon index. The resulting confidence contours of the respective observations are shown in Figure \ref{fig7:xmm_nustar_semi_simultaneous_contours} and Figure \ref{fig8:xmm_nustar_simultaneous_contours}. These results further reinforce the scenario in which the soft thermal emission arises from a geometrically thin, optically thick accretion disc, while the high-energy component is attributed to inverse Compton scattering in a hot plasma region or shock interface. Moreover, model comparison based on information criteria shows a strong preference for the inclusion of the disc component, with $\Delta \mathrm{AIC_{AB}}$ and $\Delta \mathrm{BIC_{AB}} \gtrsim 20$ for both observation (See Table \ref{tab:disk-significance}). The inferred inner disc radius remains well within both the corotation and light cylinder radii (except for the upper limit in observation N1+X3), lending strong support to the extension of accretion disc inside the magnetosphere as proposed for tMSP \citep{Campana2016,Bozzo2018,Bhattacharyya2020}.

%%%%%%%%%%%%%%%%%%%%%%%%%%%%%%%%%%%%%%%%%%%%%%%%%%%%%%%%%%%%%%%%%%%%
\begin{figure*}[ht]
        \centering
        \includegraphics[width=0.35\textwidth,angle=270]{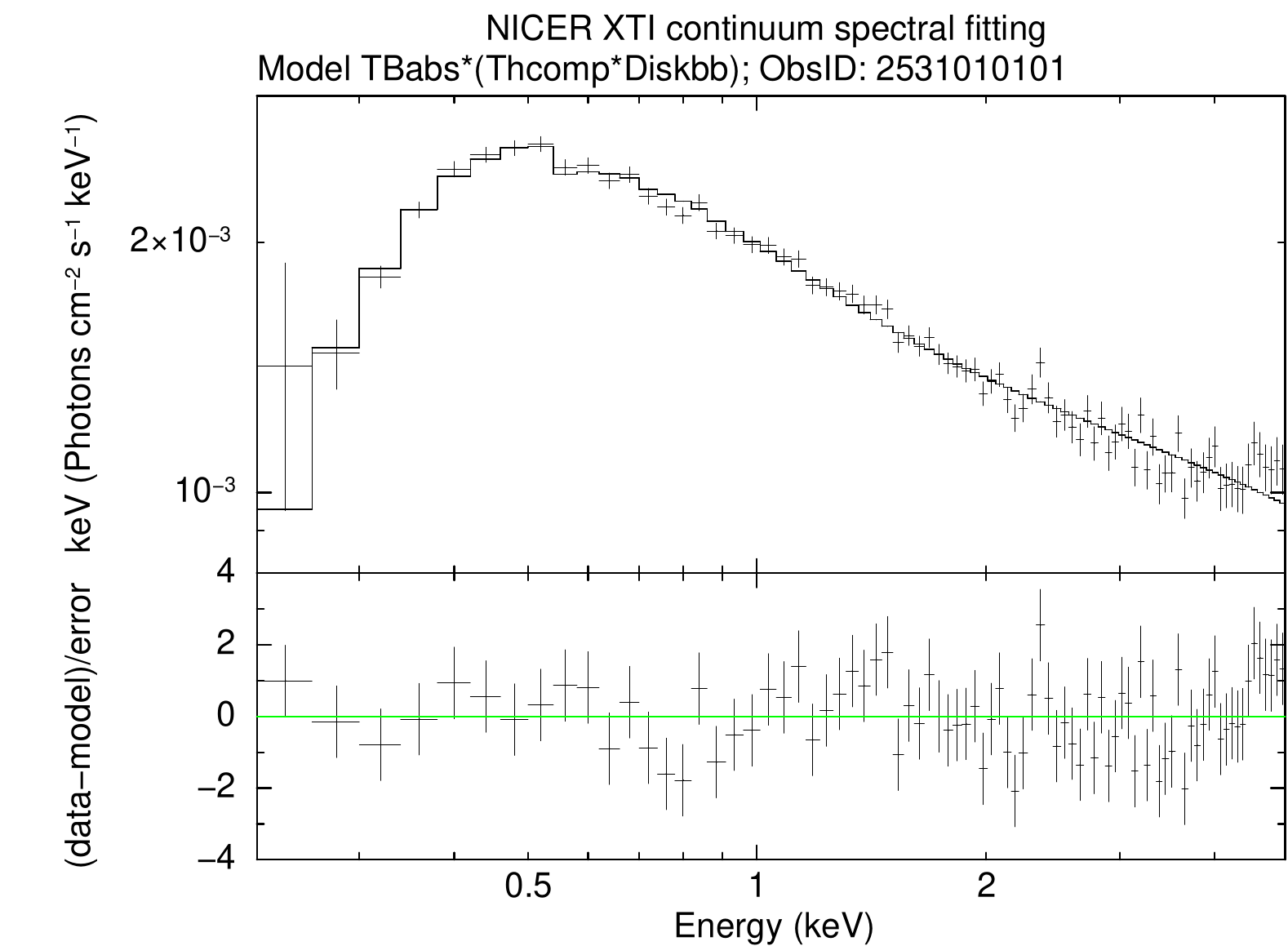}
        \hfill
        \includegraphics[width=0.35\textwidth,angle=270]{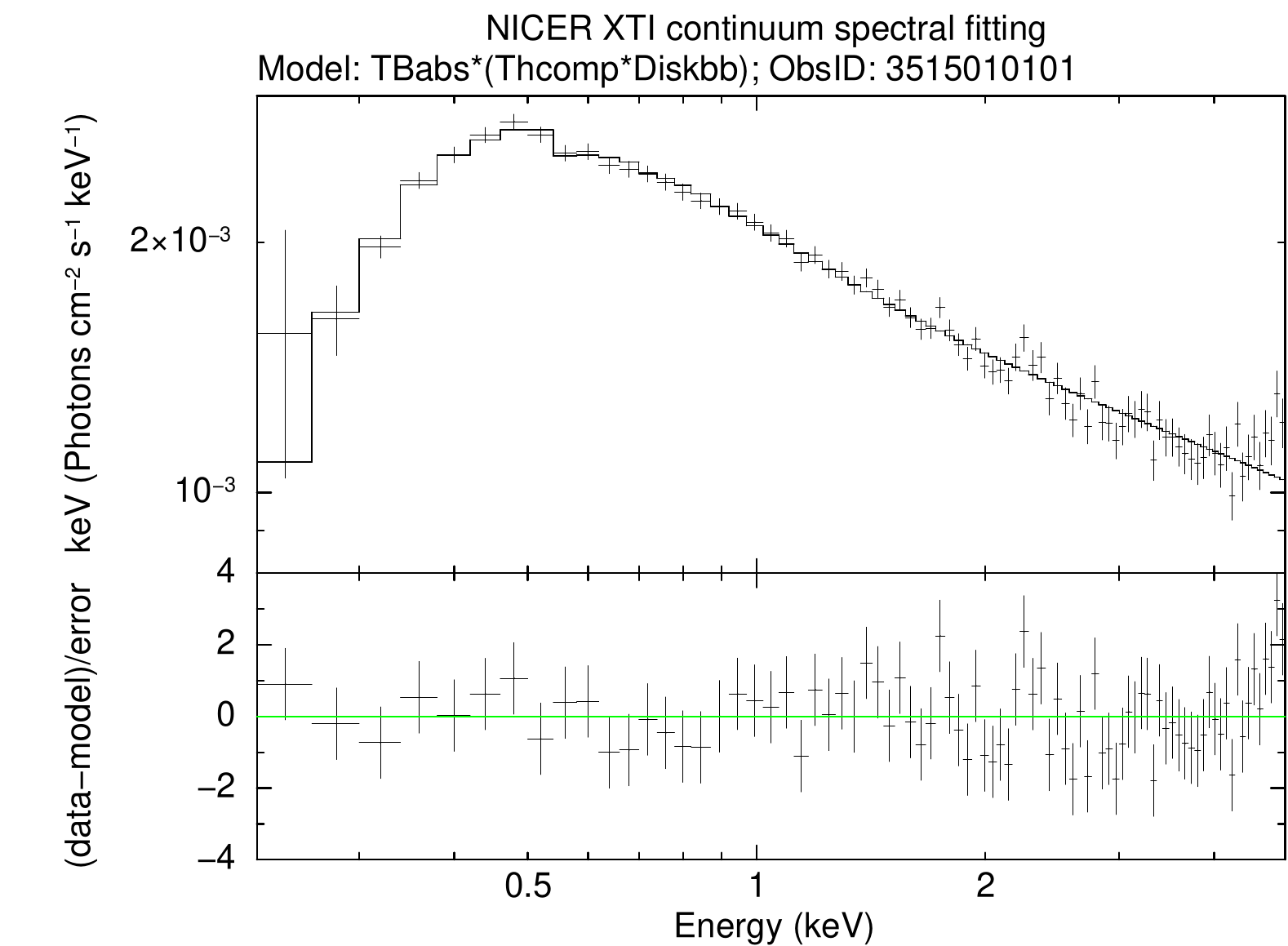}
        \caption{Best-fit \nicer{} spectra for PSR J1023+0038 during May 2019 (Left panel) and April 2020 (Right panel) observation. Due to poor signal-to-noise ratio above 5 keV, 0.2-5 keV spectra are considered for fitting.}
        \label{fig10:nicer_spectra}
\end{figure*}
%%%%%%%%%%%%%%%%%%%%%%%%%%%%%%%%%%%%%%%%%%%%%%%%%%%%%%%%%%%%%%%%%%
\begin{figure*}[ht]
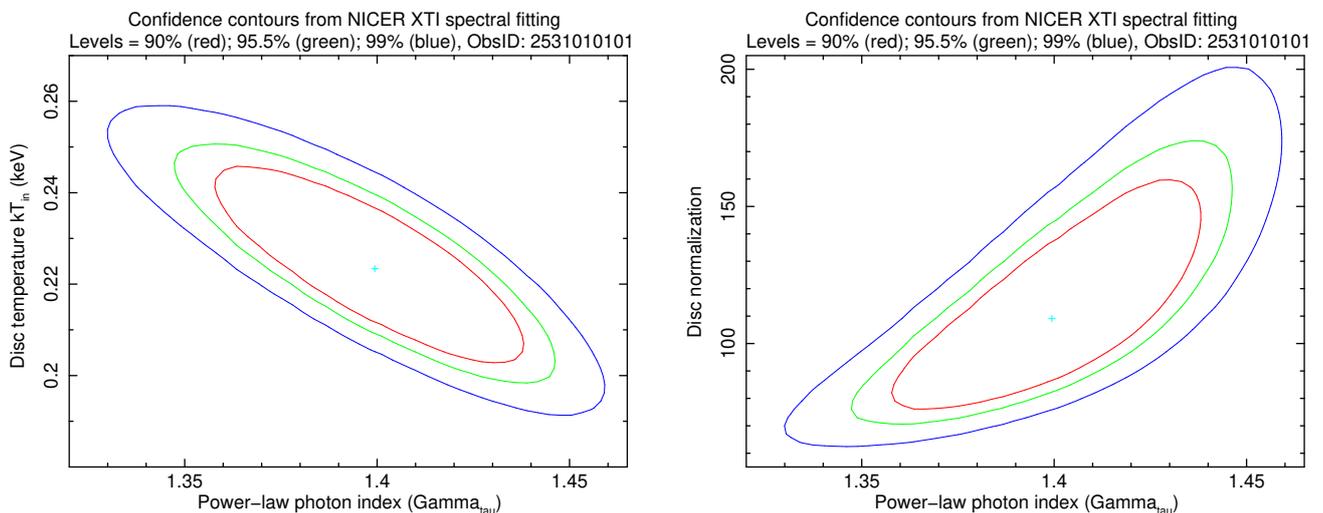

    \centering
    \includegraphics[width=0.37\textwidth,angle=270]{Fig11a-nicer_gamma_Tin_obs1_review.eps}
    \includegraphics[width=0.37\textwidth,angle=270]{Fig11b-nicer_gamma_tnorm_obs1_review.eps}
     \caption{The contours represent 90\% (red), 95.5\% (green), and 99\% (blue) confidence levels of best-fit spectral parameters for \nicer{} May 2019 observations (ObsId: 2531010101). The left and right panels show confidence contours for power-law photon index ($\Gamma_{\tau}$) versus disc temperature ($kT_{in}$), and power-law photon index ($\Gamma_{\tau}$) versus disc normalisation, respectively.  }
    \label{fig11:nicer_contours}
\end{figure*}
%%%%%%%%%%%%%%%%%%%%%%%%%%%%%%%%%%%%%%%%%%%%%%%%%%%%%%%%%%%%%%%%%%%%
%%%%%%%%%%%%%%%%%%%%%%%%%%%%%%%%%%%%%%%%%%%%%%%%%%%%%%%%%%%%%%%%%%%%%%%%%%%%%%%%%%%%%%%%%%%%%%%%%%%%%%%%%%%%%%%%%%%%%%%%%%%%%%%%%%%%%%%%%%%%%%
\begin{figure*}[ht]
    \centering
    % First row
    \includegraphics[width=0.48\textwidth]{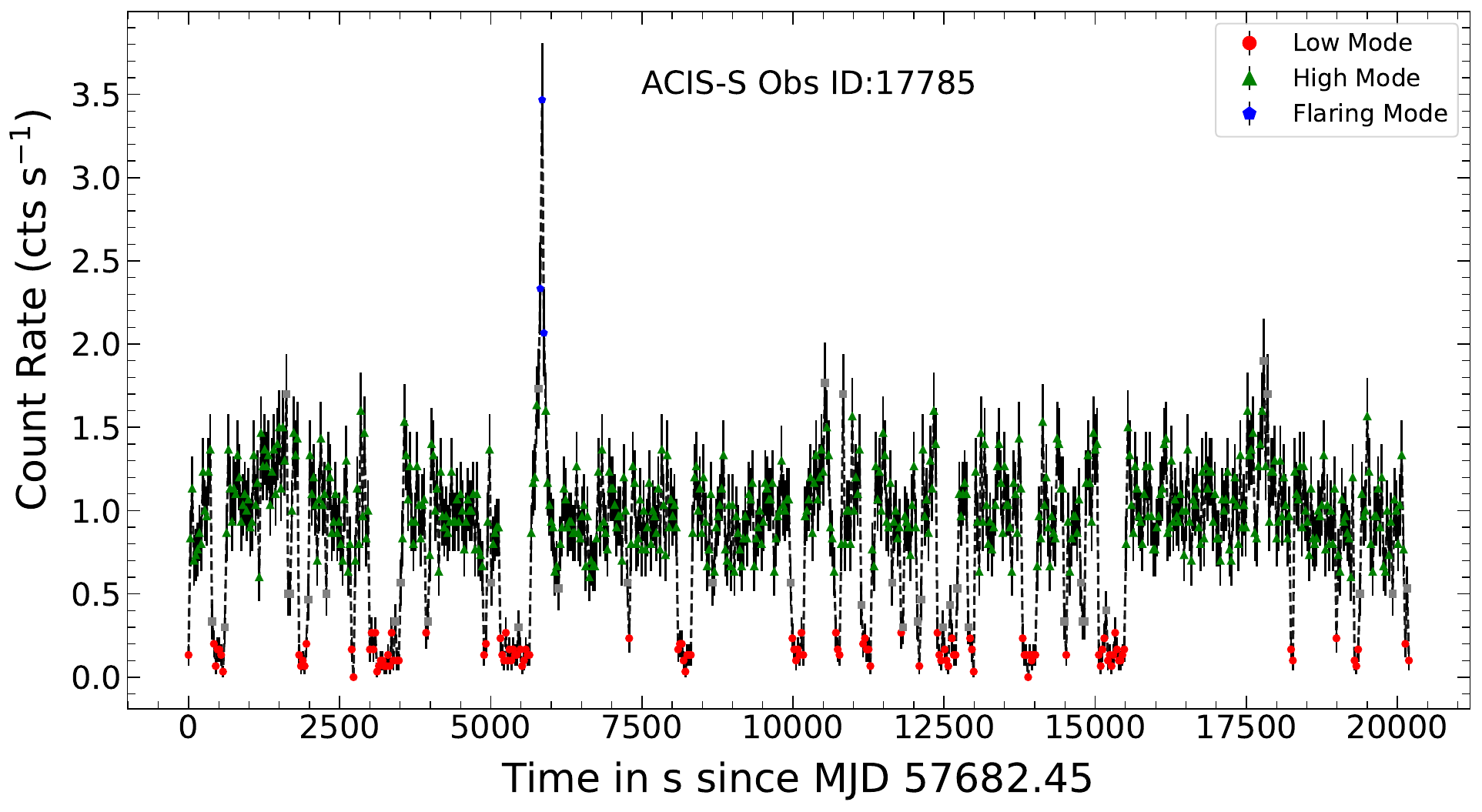}
    \includegraphics[width=0.50\textwidth]{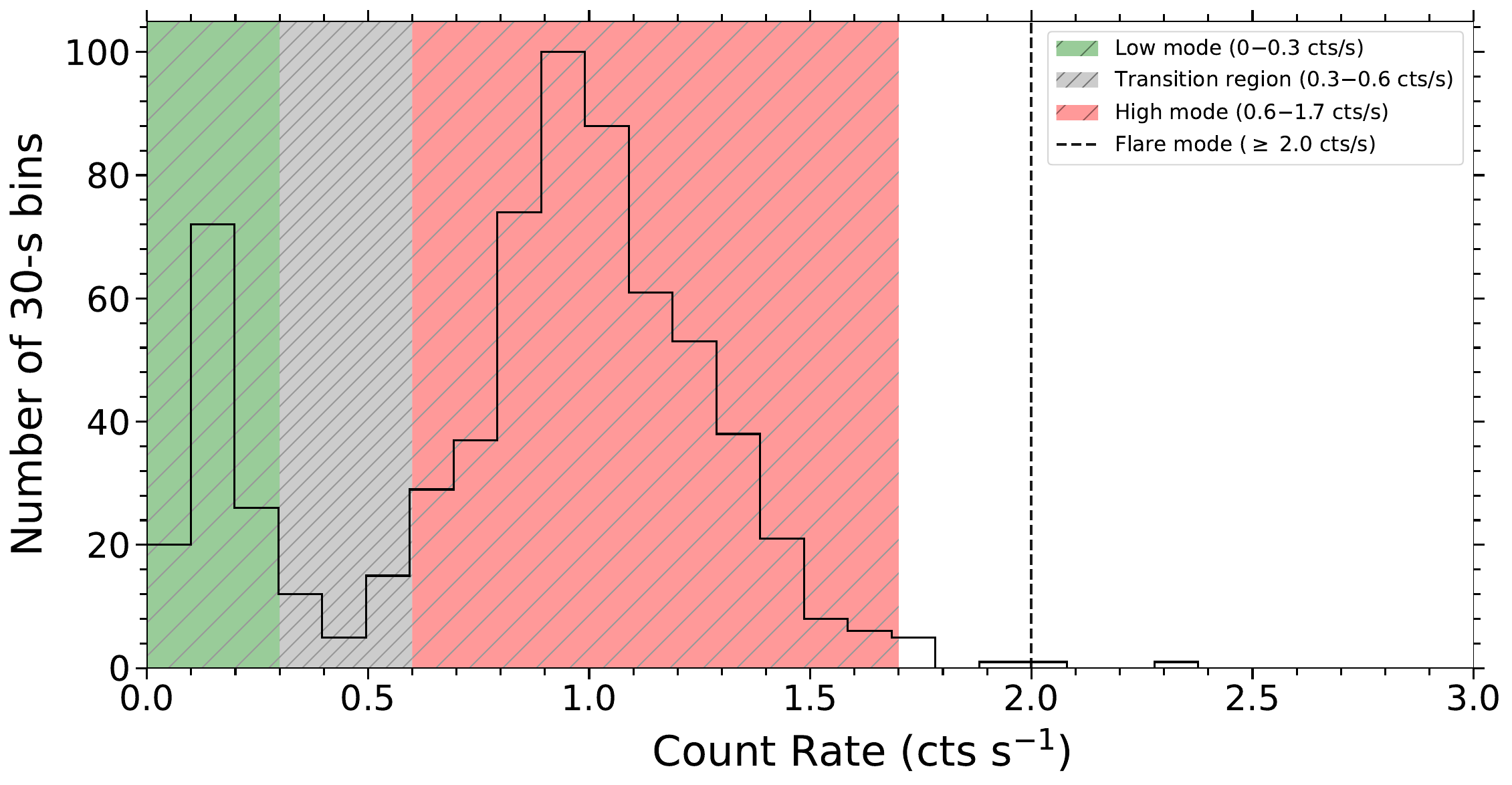}
    \caption{Left: \chandra{}/ACIS-S background-subtracted X-ray light curve of ObsId:17785 in 0.5--7.5 keV energy range binned at a time resolution of 30s. Right: Distribution of count rates produced from the 30-s binned light curves from the \chandra{} observation taken in October 2016.}
    \label{fig12:chandra_mode_selection}
\end{figure*}
%%%%%%%%%%%%%%%%%%%%%%%%%%%%%%%%%%%%%%%%%%%%%%%%%%%%%%%%%%%%%%%%%%%%%%%%%%%%%%%%%%%%%%%%%%%%%%%%%%%%%%%%%%%%%%%%%%%%%%%%%%%%%%%%%%%%%%%%%%%%%%%

\subsection{\nicer{}}
\subsubsection{X-ray variability and mode selection in \nicer{}}
\label{nicer_mode_selection}
The background-subtracted light curve of PSR J1023+0038 obtained from \nicer{} observations in the 0.5–10.0 keV energy band, with a time resolution of 10 seconds, is shown in the left panel of Figure \ref{fig9:nicer_mode_selection}. Consistent with the behaviour observed in \nus{} and \xmm{} light curves of this study, the \nicer{} light curve also shows pronounced X-ray variability characterised by distinct transitions between high, low, and flaring modes. Throughout both observations, PSR J1023+0038 predominantly resides in the high-mode, with an average count rate of $\approx \rm \,5.5 \,counts\,s^{-1}$. Rapid mode switching events are evident, with transitions to the low-mode occurring on timescales of a few tens of seconds, during which the count rate drops to an average of $\approx \rm 1.0\,counts\,s^{-1}$. In addition, bright flaring episodes are also detected, reaching peak count rates exceeding $\rm 55.0\,counts\,s^{-1}$.

To classify the different emission modes, we adopted a similar procedure used for \xmm{} and \nicer{} mode selection, employing a statistical analysis of the count rate distribution. A clear bimodal structure corresponding to the high and low modes is observed in \nicer{} observation as well, along with a distinct high-count-rate tail associated with flaring activity, as shown in the right panel of Figure \ref{fig9:nicer_mode_selection}. Based on this distribution, we define the high-mode as intervals with count rates in the range 3.5–10.0 $\rm counts\,s^{-1}$, the low-mode as those with count rates between 0.0–2.4 $\rm counts\,s^{-1}$, and the flaring-mode as intervals with count rates $\geq$ 15.0 $\rm counts\,s^{-1}$.

\subsubsection{Spectral analysis and results of \nicer{}}
To further examine the spectral characteristics of PSR J1023+0038 during its high-mode, we analysed the \nicer{} observation with the longest uninterrupted exposure. High-mode intervals were identified based on the mode-resolved light curve using count rate thresholds consistent with the classification scheme described in Section \ref{nicer_mode_selection}. High-mode spectral fitting was performed in the 0.2–5.0 keV energy range as the background becomes dominating above 5 keV. We employed the same physically motivated model described in Section \ref{xmm_spectral_analysis}, while allowing all relevant parameters to vary freely.

The \nicer{} spectra of both observations (ObsId: 2531010101, 3515010101) shown in Figure \ref{fig10:nicer_spectra}, are well described by the adopted model, yielding a statistically acceptable fit with a reduced chi-square of $\chi^2_{\nu} = 1.14$ for 79 degrees of freedom with null hypothesis probability of 19.4\% and 18.6\%, respectively. The derived photon power law index are $\Gamma_{\tau} = 1.40^{+0.04}_{-0.04}$, $1.42^{+0.04}_{-0.04}$ and the disc temperature are $0.22^{+0.02}_{-0.02}$, $0.21^{+0.02}_{-0.02}$ keV for each observation, respectively. The slightly lower photon index, compared to the values obtained from the joint \xmm{} and \nus{} analysis, is attributed to the background-dominated nature of the \nicer{} spectrum above 5 keV, which limits the constraints on the high-energy tail. In contrast, the disc temperature remains consistent within uncertainties. The inferred true inner disc radius, calculated using the equation \ref{equation2}, is $\rm R_{in}^{true} = 16.4^{+9.0}_{-4.7}$, and $18.1^{+11.9}_{-5.6}$ km, and very strong evidence for the inclusion of disk component with $\Delta \mathrm{AIC_{AB}}$ and $\Delta \mathrm{BIC_{AB}} \gtrsim 25$ for each observation. The residuals show no significant deviation across the energy band, and the fit confirms the spectral stability of the source across different epochs. To further validate this, we generated the confidence contours with 90\%, 95.5\%, and 99\% significance levels for the same set of spectral parameters described earlier in Section \ref{nustar_spectral_analysis}. The corresponding contours are presented in Figure \ref{fig11:nicer_contours}. These results support the physical scenario inferred from the broadband joint analysis and demonstrate that the \nicer{} observation independently recovers the key spectral features of PSR J1023+0038 during its high-mode.
%%%%%%%%%%%%%%%%%%%%%%%%%%%%%%%%%%%%%%%%%%%%%%%%%%%%%%%%%%%%%%%%%%%%%%%%%%%%%%%%%%%%%%%%%%%%%%%%%%%%%%%%%%%%%%%%%%%%%%%%%%%%%%%%%%%%%%%%%%%%%%%

\begin{figure*}[ht]
    \centering
    % First row
    \includegraphics[width=0.37\textwidth,angle=270]{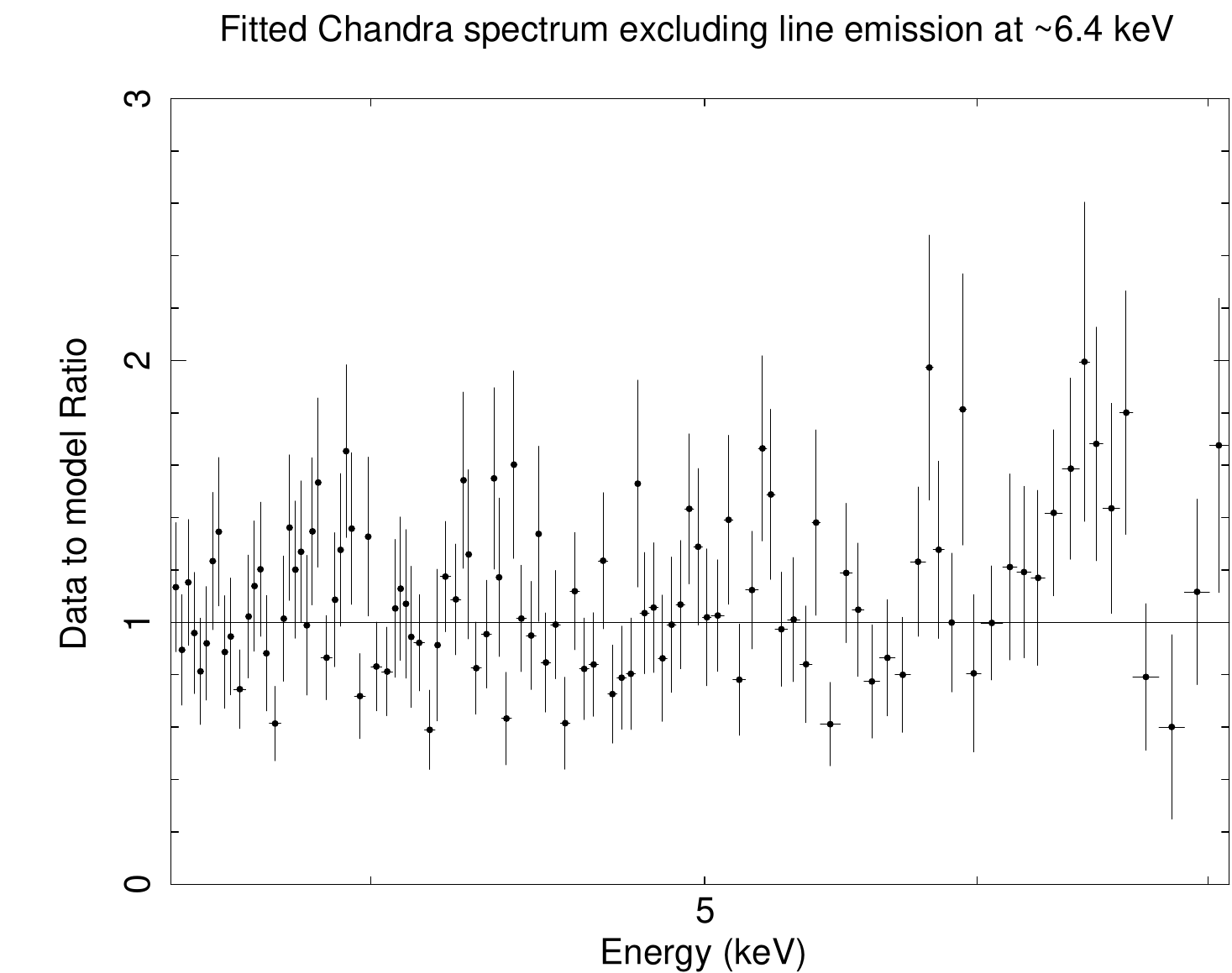}
    \hfill
    \includegraphics[width=0.37\textwidth,angle=270]{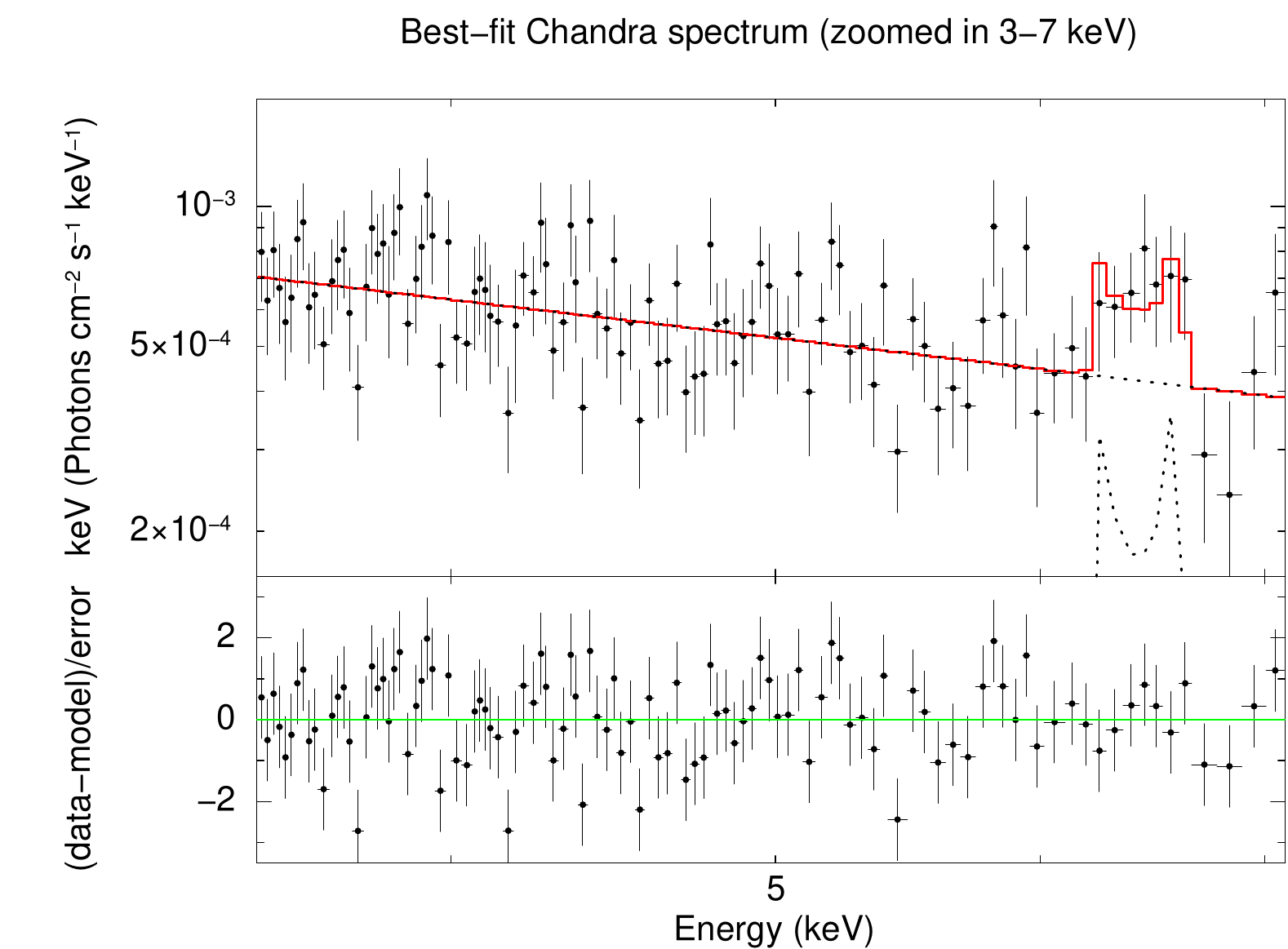}

    \includegraphics[width=0.35\textwidth,angle=270]{Fig13c-chandra_integprob-lineEN-gamma_review.eps}
    \hfill
    \includegraphics[width=0.35\textwidth,angle=270]{Fig13d-chandra_integprob-lineEN-norm_review.eps}
     \caption{Top left panel shows the residual in 3-7 keV from best-fit \chandra{} spectra without including line model at 6-7 keV. The top right panel shows the best-fit spectrum and the residual after adding the line model. Bottom left and right panel shows integrated contours from the marginal probability distribution of Fe emission line energy versus power-law photon index ($\Gamma_{\tau}$) and Fe emission line energy versus Fe line normalisation, respectively, obtained from \chandra{} spectral analysis.}
    \label{fig13:chandra_spectral_results}
\end{figure*}
%%%%%%%%%%%%%%%%%%%%%%%%%%%%%%

\subsection{\chandra{}}
\subsubsection{X-ray variability and mode selection of \chandra{}}
The background-subtracted light curve of PSR J1023+0038 from the \chandra{} observation (ObsID 17785), extracted in the 0.5–7.5 keV energy band with a time resolution of 30 seconds, is shown in the left panel of Figure \ref{fig12:chandra_mode_selection}. In contrast to other Chandra observations where the source remained in quiescence \citep{Bogdanov2011}, this particular dataset reveals clear X-ray variability characterised by rapid transitions between distinct high and low emission states \citep{Bogdanov2018}. Throughout most of the exposure, PSR J1023+0038 resides in the high-mode, with an average count rate of $\approx \rm 0.95\,counts\,s^{-1}$. Sudden drops in intensity are observed, where the source transitions into the low-mode with count rates falling to $\approx \rm 0.15\, counts\,s^{-1}$ on timescales of tens of seconds. In addition, brief flaring episodes are detected, with peak count rates exceeding 2.0 $\rm counts\,s^{-1}$ above the high-mode baseline.

To classify the emission modes in the \chandra{} data, we adopted the same statistical approach used for the \xmm{}, \nus{}, and \nicer{} observations. A bimodal distribution of count rates is clearly evident in the right panel of Figure \ref{fig12:chandra_mode_selection}, reflecting the presence of high and low-modes. Based on this distribution, we define the high-mode as intervals with count rates between 0.6--1.7 $\rm counts\,s^{-1}$ and the low-mode as those between 0.0--0.3 $\rm counts\,s^{-1}$, and the flaring-mode as intervals with count rates $\geq$ 2.0 $\rm counts\,s^{-1}$ above the high-mode mean. These findings confirm that the \chandra{} dataset captures the characteristic mode-switching behaviour of J1023, consistent with results from \nicer{}, \nus{}, and \xmm{}.

%%%%%%%%%%%%%%%%%%%%%%%%%%%%%%%%%%%%%%%%%%%%%%%%%%%%%%%%%%%%%%%%%%%%%%%%%%%%%%%%%%%%%%%%%%%%%%%%%%%%%%%%%%%%%%%%%%%%%%%%%%%%%%%%%%%%%%%%%%%%%%%%%%%%%%%%%%%%%%%%%
\begin{table*}[!ht]
\setlength{\tabcolsep}{2pt} 
\centering
\caption{Best fit parameters from spectral fitting of \xmm{}, \nus{}, \nicer{}, and {\it Chandra} observations of PSR J1023+0038.}
\label{tab:spectral_param}
\renewcommand{\arraystretch}{1.5}
\begin{tabular}{lccccccccccr} 
\hline
\hline
Component          & Parameter       &    Unit              & \xmm{}                    &    \nus{}                  &  \nus{}                  &        \nicer{}           &      \nicer{}            &      Chandra    \\
                   &                 &                      & (X1+X2+X3)                &    (N1+X3)                 &  (N2+X4)                 &        2531010101         &    3515010101            &      17785      \\
\hline 

TBabs              &  $ n_{\rm H}$   & $10^{20}$ cm$^{-2}$  & $2.915_{-0.003}^{+0.004}$ & $4.19_{-0.01}^{+0.01}$     & $3.80_{-0.01}^{+0.01}$   & $2.23_{-0.01}^{+0.01}$    & $2.04_{-0.01}^{+0.01}$   & $5.82_{-0.89}^{+0.95}$ \\ 

Diskbb             & kT$_{\rm disc}$ &         keV          & $0.21_{-0.01}^{+0.01}$    & $0.18_{-0.01}^{+0.01}$     & $0.20_{-0.01}^{+0.01}$   & $0.22_{-0.02}^{+0.02}$    & $0.21_{-0.02}^{+0.02}$   & $0.12_{-0.02}^{+0.05}$\\
                   & $N_{disc}$$^\ast$ &                    & $115_{-20}^{+25}$         & $232_{-61}^{+65}$          & $140_{-23}^{+45}$        & $109_{-33}^{+50}$         & $133_{-44}^{+74}$        & $<1026$\\  

Thcomp             & $\Gamma_{\tau}$ &                      & $1.66_{-0.01}^{+0.01}$    & $1.69_{-0.01}^{+0.01}$     & $1.65_{-0.01}^{+0.02}$   &  $1.40_{-0.04}^{+0.04}$   & $1.42_{-0.04}^{+0.04}$    &  $1.88_{-0.04}^{+0.03}$\\
                   & kT$_{e}$        & keV                  & $113_{-14}^{+13}$         &  $>96$                     & $>109$                   &  $113^\star$              & $113^\star$               &  $>33$ \\
                   & cov\_frac $(f)$ &                      & $0.91^\star$              & $0.91^\star$              & $0.91^\star$                  & $0.86_{-0.05}^{+0.04}$    &  $0.91_{-0.05}^{+0.06}$   &   $0.91^\star$                   \\

diskline           & Line            &   keV                &       --                  &    --                    & --                      & --                      &    --                   & $6.45_{-0.04}^{+0.04}$\\
                   & R$_{in}$        &   (R$_{g}$)          &       --                  &    --                    & --                      & --                      &     --                  & $<21$              \\
                   & Norm            &   (x$10^{-5}$)       &       --                  &    --                    & --                      & --                      &     --                  &$1.83_{-0.57}^{+0.91}$ \\
constant           & $\rm C_{MOS}$   &                      &       1.0              &    1.0                   & 1.0                     & --                      &       --                &    --                 \\
                   & $\rm C_{PN}$    &                      &$1.017_{-0.004}^{+0.004}$ & $0.99_{-0.01}^{+0.01}$ & $1.08_{-0.01}^{+0.01}$  & --                      &      --                 &    --                  \\
                   & $\rm C_{FPMA}$  &                      &         --             & $1.12_{-0.02}^{+0.02}$   & $1.12_{-0.04}^{+0.04}$  & --                      &             --          &    --                   \\
                   & $\rm C_{FPMB}$  &                      &         --             & $1.11_{-0.02}^{+0.02}$   & $1.13_{-0.04}^{+0.04}$  & --                      &       --                &    --                 \\                   

\hline
F$_{0.3-10\,keV}$&           & $10^{-12}$ ergs/s/cm$^{-2}$  & $12.11_{-0.02}^{+0.02}$ & $12.43_{-0.06}^{+0.06}$& $11.55_{-0.07}^{+0.07}$& $17.10_{-0.10}^{+0.10}$ & $18.10_{-0.11}^{+0.11}$ &  $11.23_{-0.33}^{+0.37}$\\
\hline
Test statistic     & $\chi^2$/dof    &                      &        2183/2057       &    1619/1586              & 1192/1152                & 90/79                 &      90/79           &      299/305\\

\hline
\end{tabular}
\begin{flushleft}
\textbf{Note:} Errors in each parameter are quoted with 90\% confidence intervals ($\Delta\chi^2=2.706$). Upper and lower limits are provided with similar confidence for a few parameters for which errors are unconstrained. \\
$\ast$: The disc normalisation is defined as $\rm N_{disc} =(r_{in}/D_{10})^2*\cos \theta $, where $\rm r_{in}$ is the apparent inner radius of the accretion disc in km, $D_{10}$ is the distance to the source in units of 10~kpc, and $\theta$ is the inclination angle of the disc in degrees.  \\
$\star$: The parameter is kept fixed. \\
F$_{0.3-10\,keV}$: Absorbed X-ray flux in 0.3-10.0 keV energy range.
\end{flushleft}
\end{table*}
%%%%%%%%%%%%%%%%%%%%%%%%%%%%%%%%%%%%%%%%%%%%%%%%%%%%%%%%%%%%%%%%%%%%%%%%%%%%%%%%%%%%%%%%%%%%%%%%%%%%%%%%%%%%%%%%%%%%%%%%%%%%%%%%%%%%%%%%%%%%%%%%%%%%%%%%%%%%%%%%%%%%
%%%%%%%%%%%%%%%%%%%%%%%%%%%%%%%%%%%%%%%%%%%%%%%%%%%%%%%%%%%%%%%%%%%%%%%%%%%%%%%%%%%%%%%%%%%%%%%%%%%%%%%%%%%%%%%%%%%%%%%%
\begin{table*}[!ht]
    \centering
    \caption{Model A: {\textsc TBabs* (Thcomp$\otimes$Diskbb)}, Model B: {\textsc TBabs* (Thcomp$\otimes$bbodyrad)}, Model C: {\textsc TBabs*zxipcf*(Nsatmos+Powerlaw+Powerlaw)}. Model comparison is based on the Akaike Information Criterion (AIC) and the Bayesian Information Criterion (BIC). $\rm R_{in}^{true}$ was calculated by considering a color correction factor of $1.85_{-0.15}^{+0.15}$, general relativity correction factor of 0.412, distance of $1.37_{-0.04}^{+0.04}$ kpc, inclination of $42_{-2}^{+2}$ degrees, and normalisation factor of 7.}
    \renewcommand{\arraystretch}{1.5} 
    \setlength{\tabcolsep}{3.5pt} 
    \begin{tabular}{ccccccccccr}
    \hline
    \hline
        Observatory         &  Observation    & Model A               & Model B             & Model C                &   $\Delta BIC_{AB}$  & $\Delta BIC_{AC}$ &     $N_{disc}$        & $\rm R_{\rm in}^{true}\,({\rm km})$ \\
                            &    (ObsId)      & $\chi^2/\nu, (k)$     & $\chi^2/\nu, (k)$   & $\chi^2/\nu, (k)$      &                      &                 &  (3$\sigma$ error)    & (3$\sigma$ error)     \\
    \hline  
        \xmm{}              & X1+X2+X3        &  2183/2057, (6)       &  2302/2057, (6)     & 2300/2057, (6)         &     119              &    117          &  $115_{-35}^{+36}$    & $16.8_{-3.8}^{+3.8}$   \\
        \multirow{2}{*}{\raisebox{-1.0ex}{\shortstack{\nus{} \\ + \\ \xmm{}}}}
                            & N1+X3           &  1619/1586, (8)       &  1641/1586, (8)     & 1647/1586, (8)         &     22               &    28           &  $232_{-107}^{+115}$  & $23.2_{-6.6}^{+6.9}$   \\
                            & N2+X4           &  1192/1152, (8)       &  1225/1152, (8)     & 1229/1152, (8)         &     33               &    37           &  $140_{-70}^{+87}$    & $18.6_{-5.6}^{+6.5}$   \\
        \nicer{}            & 2531010101      &  90/79, (5)           &  117/79, (5)        & 83/79, (5)             &     27               &    -7           &  $109_{-52}^{+114}$   & $16.4_{-4.7}^{+9.0}$   \\
        \nicer{}            & 3515010101      &  90/79, (5)           &  123/79, (5)        & 84/79, (5)             &     33               &    -6           &  $133_{-69}^{+169}$   & $18.1_{-5.6}^{+11.9}$   \\
       
    \hline
    \end{tabular}
     \begin{flushleft}
    \textbf{Note.}: Model A, Model B, and Model C are evaluated using:
    \begin{align*}
        \mathrm{AIC} &= \chi^2 + 2k \\
        \mathrm{BIC} &= \chi^2 + k \cdot \ln(N), \quad \text{where } N = \nu + k \\
        \Delta \mathrm{AIC_{AB}} &= \mathrm{AIC}_{\text{Model B}} - \mathrm{AIC}_{\text{Model A}}, \,\Delta \mathrm{BIC_{AB}} = \mathrm{BIC}_{\text{Model B}} - \mathrm{BIC}_{\text{Model A}}\\
        \Delta \mathrm{AIC_{AC}} &= \mathrm{AIC}_{\text{Model C}} - \mathrm{AIC}_{\text{Model A}}, \,\Delta \mathrm{BIC_{AC}} = \mathrm{BIC}_{\text{Model C}} - \mathrm{BIC}_{\text{Model A}}
    \end{align*}
    Here, \(k\) denotes the number of free parameters, and \(\nu\) the number of degrees of freedom. Estimated $\Delta AIC_{AB}$ and $\Delta AIC_{AC}$ are identical to $\Delta BIC_{AB}$ and $\Delta BIC_{AC}$, and $\Delta AIC_{AB,AC}, \, \Delta BIC_{AB,AC} > 10$ indicates very strong evidence in favor of Model A \citep{Shi2012}.
    \end{flushleft}
    \label{tab:disk-significance}
\end{table*}
%%%%%%%%%%%%%%%%%%%%%%%%%%%%%%%%%%%%%%%%%%%%%%%%%%%%%%%%%%%%%%%%%%%%%%%%%%%%%%%%%%%%%%%%%%%%%%%%%%%%%%%%%%%%%%%%%%%%%%%%

\subsubsection{Spectral analysis and results of \chandra{}}
Following previous approaches, we used a combination of \textsc{diskbb} and \textsc{thcomp} models to fit the continuum due to thermal and Comptonisation components. With the continuum model, we observed a strong, systematic residual around $\sim$6-7 keV. The residual is shown in the top left panel of Figure \ref{fig13:chandra_spectral_results}. Such a residual is usually due to the presence of the Fe k$\alpha$ line complex.
We added the {\textsc diskline} model to the continuum to fit the line complex. The fit improves significantly with the $\chi^2$/dof = 299/305 (the $\chi^2$/dof without the {\textsc diskline} model is 313/307), provides an F-test probability of $9.3\times10^{-4}$ and significance of $3.1\sigma$ for \textsc{diskline} component. The line is observed at 6.45$^{+0.04}_{-0.04}$ keV with the 90\% upper limit of the inner disc radius of 21 R$_g$. The fitted model and the residual are shown in the top right panel of Figure \ref{fig13:chandra_spectral_results}. We also fitted the Fe line complex with a Gaussian model that gives a similar test statistic of $\chi^2$/dof = 297/304 with Fe line energy of 6.44$^{+0.08}_{-0.10}$ keV and line width of $0.12^{+0.18}_{-0.08}$ keV.
To further assess the significance of the detection of the Fe line and as an independent verification of our results from the $\chi^2$ minimisation method, we employ a Markov Chain Monte Carlo (MCMC) simulation approach. Given that X-ray spectral counts typically follow a Poisson distribution, we replace the standard $\chi^2$ fitting statistic with {\textsc pgstat} in {\textsc XSpec}, which is specifically designed for Poisson-distributed source counts and assumes Gaussian-distributed background counts. The profile likelihood formulation of {\textsc pgstat} is analogous to that used in the derivation of the C-statistic likelihood. Using this updated fit statistic, we run MCMC chains of 1 $\times$ 10$^5$ steps, initiating the process from a randomly perturbed position relative to the best-fit parameters and discarding the initial 10000 steps as burn-in. The proposal distribution at each Monte Carlo step is assumed to be Gaussian, scaled by a factor of 0.001. For the MCMC sampling, we adopt the Goodman–Weare algorithm, utilising 10 walkers. Contours between the line energy and line normalisation are shown in the bottom panels of Figure \ref{fig13:chandra_spectral_results} with 90\%, 95\%, and 99\% significances. Models for relativistic line profiles like {\textsc laor} do not provide a significantly better fit.

For all observations, the true inner disc radius is estimated using equations \ref{equation2}, and lies within the range of 12--30 km with a significance of 3$\sigma$. For \chandra{}, the inner disc radius was not estimated as the parameters were unconstrained within 3$\sigma$ confidence.

%%%%%%%%%%%%%%%%%%%%%%%%%%%%%%%%%%%%%%%%%%%%%%%%%%%%%%%%%%%%%%%%%%%%%%%%%%%%%%%%%%%%%%%%%%%%%%%%%%%%%%%%%%%%%%%%%%%%%%%%%%%%%%%%%%%%%%%%%%%%%%%%%%%%%
\subsection{Spectral modelling comparison with previous studies}
We have compared our spectral model, \textsc{TBabs* (Thcomp$\otimes$Diskbb)}, with that proposed by \citet{Campana2019}, \textsc{tbabs*zxipcf*(nsatmos+powerlaw+powerlaw)}, to evaluate the performance in describing the observed spectra. 
We modelled our observed high-mode spectrum of PSR J1023+0038 using the multi-component approach proposed by \citet{Campana2019}, which considers a hotspot emission component from the neutron star surface, represented by a neutron star atmosphere model \citep[\textsc{nsatmos};][]{Heinke2006} with free radius, assuming a neutron star mass of $1.4 M_{\odot}$, radius of 10 km, and a distance of 1.37 kpc \citep{Deller2012}. For which, the hot emission region was fixed at a temperature of $\log (T/{K})=5.85$, with the emission fraction of the neutron star surface set to $f=0.16$. The magnetospheric emission was constrained using a power-law component with the photon index ($\Gamma_{mag}$) fixed at 1.06, while shock emission from high-energy photons produced at the shock front was described with an additional power-law ($\Gamma_{shock}$). To account for absorption, a partially ionised absorber was included using the \textsc{zxipcf} model \citep{Reeves2008}, with the column density of the hot absorber fixed at $\rm 23.4\times10^{22}\,cm^{-2}$. In addition, the interstellar absorption along the line of sight, modelled with \textsc{tbabs} \citep{Wilms2000} and fixed to a column density of $\rm 5.0\times10^{20}{cm^{-2}}$ during the spectral fitting \citep[See section 4 of][]{Campana2019}. 

In comparison, our model describes the soft thermal emission with a multi-temperature disc blackbody \citep[\textsc{diskbb};][]{Mitsuda1984,Makishima1986}, while the hard X-ray emission is modeled with a physically motivated Comptonisation component \citep[\textsc{thcomp};][]{Zdziarski2020}, along with the \textsc{tbabs} model implemented for interstellar absorption.

To statistically assess the goodness of spectral fit and model preference, we computed information criteria AIC and BIC for both models \citep{Akaike1974,Schwarz1978,Liddle2007}. The model comparison based on these criteria is summarised in Table~\ref{tab:disk-significance}, showing a strong preference \citep[i.e. $\Delta$AIC and $\Delta$BIC $\gtrsim10$;][]{Shi2012} and a significantly better fit for our model (Model A) over \citet{Campana2019} model (Model C) for the \xmm{} and joint \xmm{}+\nus{} observations. For the \xmm{} observation (X1+X2+X3), $\rm \Delta AIC_{AC}$ and $\rm \Delta BIC_{AC}$ both exceed 115, which provides a strong evidence in favour of Model A. Similarly, the N1+X3 and N2+X4 joint \xmm{}+\nus{} observations yield significant positive differences ($\rm \Delta AIC_{AC} \approx  \rm \Delta BIC_{AC} \sim 28-37$), again indicating very strong support for Model A. However, for the \nicer{} observations, both $\rm \Delta AIC_{AC}$ and $\rm \Delta BIC_{AC}$ are negative. These results show that the preference for Model A is robust for high-quality \xmm{} and joint \xmm{}+\nus{} datasets, whereas \nicer{}’s more limited spectral coverage and resolution are results in a weak preference for Model A. Overall, these findings establish that our simpler model effectively constrains the dominant thermal and Comptonised components of the spectrum without requiring additional absorption or multiple non-thermal components, supporting the presence of an accretion disc inside the $r_{co}$ during the high-mode of the active state of PSR J1023+0038.
%%%%%%%%%%%%%%%%%%%%%%%%%%%%%%%%%%%%%%%%%%%%%%%%%%%%%%%%%%%%%%
\section{Discussion and conclusion}\label{Discussion}

The extent to which the accretion disc in the high-mode of transitional millisecond pulsars (tMSPs) penetrates the magnetosphere and reaches near the neutron star surface remains an open question. We investigate this in the prototypical tMSP PSR J1023+0038 by isolating the high-mode intervals and performing a detailed, multi-epoch spectral analysis. Our dataset includes the longest joint observation from \xmm{} EPIC-PN, MOS1, and MOS2; simultaneous \xmm{}+\nus{} observations; and individual exposures from \nicer{} and \chandra{}. 

Using the longest available exposures, by combining three \xmm{} exposures (approximately 202 ksec of high-mode data out of a combined exposure of about 364 ksec), we performed a self-consistent spectral analysis. From this, we constrained the inner disc radius to be 16.8$\pm$3.8 km, with a significance of at least 3$\sigma$.
This measurement is consistent, within 3$\sigma$ uncertainties, with the best-fit inner disc radius values (12-30 km) obtained from other observatories such as \nicer{}, as well as from joint observations with \xmm{} and \nus{}. Additionally, we detect an Fe K$\alpha$ emission line at 6.45 keV in the \chandra{} spectrum with 99\% confidence, estimated by Markov Chain Monte Carlo simulations with the 3$\sigma$ upper limit of the inner disc radius of 21 R$_g$. These results strongly suggest that, during the high-mode, the accretion disc in PSR J1023+0038 extends inside the light-cylinder radius of the neutron star, indicating a significant presence of disc component. The narrow Fe line feature observed during the \chandra{} observation, attributed to reflection, also provides independent evidence for the presence of the inner accretion disc during high mode.

Our results have broader implications for the torque dynamics and potential gravitational wave (GW) emission mechanisms in transitional millisecond pulsars as proposed in \citet{Bhattacharyya2020}. A rotating neutron star with a non-zero ellipticity, corresponding to a non-axisymmetric mass distribution, is theoretically expected to emit continuous gravitational waves. Although such emissions have not yet been directly detected, their influence can be inferred indirectly through contributions to the pulsar’s spin-down torque.
PSR J1023+0038 is uniquely suited for this analysis, as it remains the only known system with measured spin-down rates in both its accreting and non-accreting states. By applying a torque budget formalism that incorporates electromagnetic dipole radiation, GW emission, and accretion torques under the assumption that $\Gamma$-ray emission originates from the pulsar magnetosphere in both states, \citet{Bhattacharyya2020} inferred the presence of a persistent ellipticity, which should generate continuous gravitational waves.
Notably, the same framework also provides a coherent explanation for several other key observational phenomena. It predicts the inward extension of the accretion disc into the neutron star’s magnetosphere during the high-mode, consistent with our spectral results. Moreover, it, supported by our spectral results, naturally accounts for the observed coherent X-ray pulsations during the accreting state via standard channelled accretion models, without invoking exotic mechanisms.

\section*{Acknowledgements}
We thank the referee for valuable comments and suggestions that helped us improve the manuscript. VJ thanks the fellowships from the Ministry of Education, Government of India, and the Department of Physics, IIT Hyderabad, for providing support.
SB acknowledges financial support by the Fulbright-Nehru Academic \& Professional Excellence Award (Research), sponsored by the U.S. Department of State and the United States-India Educational Foundation (grant number: 3062/F-N APE/2024; program number: G-1-00005).

%\bibliographystyle{elsarticle-harv} 
%\bibliography{PSRJ1023}

\end{document}